\documentclass[journal,twoside]{IEEEtran}

\usepackage[utf8]{inputenc}
\usepackage[T1]{fontenc}
\usepackage[english]{babel}
\usepackage{fontenc}
\usepackage{graphicx}
\usepackage{listings}
\usepackage{amsmath}
\usepackage{eso-pic}
\usepackage{float}
\usepackage{picinpar}
\usepackage{color}
\usepackage{environ}
\usepackage{tikz}
\usetikzlibrary{trees}
\usetikzlibrary{snakes}
\usetikzlibrary{calc,matrix}
\usepackage{multicol}
\usepackage{type1cm}
\usepackage{xcolor}
\usepackage{enumitem}
\usepackage{longtable}
\usepackage{verbatim}
\usepackage{transparent}
\usepackage{cite}
\usepackage[hyphens]{url}
\usepackage{hyperref}
\usepackage{array}

\hypersetup{
    pdftoolbar=true,        
    pdfmenubar=true,        
    pdffitwindow=false,     
    pdfstartview={FitH},    
    pdftitle={Web Tracking: Mechanisms, Implications, and Defenses},    
    pdfauthor={Tomasz Bujlow},     
    pdfsubject={web tracking},   
    pdfcreator={PDFLateX},   
    pdfproducer={PDFLateX}, 
    pdfkeywords={web tracking} {tracking mechanisms} {tracking implications} {defenses against tracking} {user identification} {tracking discovery} {future of tracking}, 
    pdfnewwindow=true,      
    colorlinks=true,       
    linkcolor=black,          
    citecolor=black,        
    filecolor=black,      
    urlcolor=black,          
    breaklinks=true         
}

\makeatletter
\newcount\dirtree@lvl
\newcount\dirtree@plvl
\newcount\dirtree@clvl
\def\dirtree@growth{%
  \ifnum\tikznumberofcurrentchild=1\relax
  \global\advance\dirtree@plvl by 1
  \expandafter\xdef\csname dirtree@p@\the\dirtree@plvl\endcsname{\the\dirtree@lvl}
  \fi
  \global\advance\dirtree@lvl by 1\relax
  \dirtree@clvl=\dirtree@lvl
  \advance\dirtree@clvl by -\csname dirtree@p@\the\dirtree@plvl\endcsname
  \pgf@xa=1cm\relax
  \pgf@ya=-1cm\relax
  \pgf@ya=\dirtree@clvl\pgf@ya
  \pgftransformshift{\pgfqpoint{\the\pgf@xa}{\the\pgf@ya}}%
  \ifnum\tikznumberofcurrentchild=\tikznumberofchildren
  \global\advance\dirtree@plvl by -1
  \fi
}

\tikzset{
  dirtree/.style={
    growth function=\dirtree@growth,
    every node/.style={
    anchor=north
    shape = rectangle,
    rounded corners,
    align = left, anchor = west,
    draw = black,
    font={\itshape},
    },
    every child node/.style={anchor=west},
    edge from parent path={(\tikzparentnode\tikzparentanchor) |- (\tikzchildnode\tikzchildanchor)}
  }
}
\makeatother

\makeatletter
\let\matamp=&
\catcode`\&=13
\makeatletter
\def&{\iftikz@is@matrix
  \pgfmatrixnextcell
  \else
  \matamp
  \fi}
\makeatother

\newcounter{lines}
\def\endlr{\stepcounter{lines}\\}

\newcounter{vtml}
\newif\ifvtimelinetitle
\newif\ifvtimebottomline
\tikzset{description/.style={
  column 2/.append style={#1}
 },
 timeline color/.store in=\vtmlcolor,
 timeline color=red!80!black,
 timeline color st/.style={fill=\vtmlcolor,draw=\vtmlcolor},
 use timeline header/.is if=vtimelinetitle,
 use timeline header=false,
 add bottom line/.is if=vtimebottomline,
 add bottom line=false,
 timeline title/.store in=\vtimelinetitle,
 timeline title={},
 line offset/.store in=\lineoffset,
 line offset=4pt,
}

\NewEnviron{vtimeline}[1][]{%
\setcounter{lines}{1}%
\stepcounter{vtml}%
\begin{tikzpicture}[column 1/.style={anchor=east},
 column 2/.style={anchor=west},
 text depth=0pt,text height=1ex,
 row sep=1ex,
 column sep=1em,
 #1
]
\matrix(vtimeline\thevtml)[matrix of nodes]{\BODY};
\pgfmathtruncatemacro\endmtx{\thelines-1}
\path[timeline color st] 
($(vtimeline\thevtml-1-1.north east)!0.5!(vtimeline\thevtml-1-2.north west)$)--
($(vtimeline\thevtml-\endmtx-1.south east)!0.5!(vtimeline\thevtml-\endmtx-2.south west)$);
\foreach \x in {1,...,\endmtx}{
 \node[circle,timeline color st, inner sep=0.15pt, draw=white, thick] 
 (vtimeline\thevtml-c-\x) at 
 ($(vtimeline\thevtml-\x-1.east)!0.5!(vtimeline\thevtml-\x-2.west)$){};
 \draw[timeline color st](vtimeline\thevtml-c-\x.west)--++(-3pt,0);
 }
 \ifvtimelinetitle%
  \draw[timeline color st]([yshift=\lineoffset]vtimeline\thevtml.north west)--
  ([yshift=\lineoffset]vtimeline\thevtml.north east);
  \node[anchor=west,yshift=16pt,font=\large]
   at (vtimeline\thevtml-1-1.north west) 
   {\textsc{Timeline \thevtml}: \textit{\vtimelinetitle}};
 \else%
  \relax%
 \fi%
 \ifvtimebottomline%
   \draw[timeline color st]([yshift=-\lineoffset]vtimeline\thevtml.south west)--
  ([yshift=-\lineoffset]vtimeline\thevtml.south east);
 \else%
   \relax%
 \fi%
\end{tikzpicture}
}

\begin{document}


\title{Web Tracking: Mechanisms, Implications, and Defenses}

\author{Tomasz~Bujlow,~\IEEEmembership{Member,~IEEE,}
        Valentín~Carela-Español,
        Josep~Solé-Pareta,
        and~Pere~Barlet-Ros
\thanks{The authors are with the Broadband Communications Research Group, Department of Computer Architecture,
Universitat Politècnica de Catalunya, Barcelona, 08034, Spain.}
\thanks{E-mails: tomasz@bujlow.com (T.~Bujlow), vcarela@ac.upc.edu (V.~Carela-Español), pareta@ac.upc.edu (J.~Solé-Pareta), pbarlet@ac.upc.edu (P.~Barlet-Ros).}}

\markboth{arXiv.org Digital Library}%
{Bujlow \MakeLowercase{\textit{et al.}}: Web Tracking: Mechanisms, Implications, and Defenses}

\maketitle

\begin{abstract}
This articles surveys the existing literature on the methods currently used by web services to track the user online as well as their purposes, implications, and possible user's defenses. A significant majority of reviewed articles and web resources are from years 2012 -- 2014. Privacy seems to be the Achilles' heel of today's web. Web services make continuous efforts to obtain as much information as they can about the things we search, the sites we visit, the people with who we contact, and the products we buy. Tracking is usually performed for commercial purposes. We present 5 main groups of methods used for user tracking, which are based on sessions, client storage, client cache, fingerprinting, or yet other approaches. A special focus is placed on mechanisms that use web caches, operational caches, and fingerprinting, as they are usually very rich in terms of using various creative methodologies. We also show how the users can be identified on the web and associated with their real names, e-mail addresses, phone numbers, or even street addresses. We show why tracking is being used and its possible implications for the users. For example, we describe recent cases of price discrimination, assessing financial credibility, determining insurance coverage, government surveillance, and identity theft. For each of the tracking methods, we present possible defenses. Some of them are specific to a particular tracking approach, while others are more universal (block more than one threat) and they are discussed separately. Apart from describing the methods and tools used for keeping the personal data away from being tracked, we also present several tools that were used for research purposes -- their main goal is to discover how and by which entity the users are being tracked on their desktop computers or smartphones, provide this information to the users, and visualize it in an accessible and easy to follow way. Finally, we present the currently proposed future approaches to track the user and show that they can potentially pose significant threats to the users' privacy.
\end{abstract}


\section{Introduction}
\label{sec:introduction}

It is widely known that service providers (e.g., YouTube, elPais), content providers (e.g., Google, Facebook, and Amazon), and other third parties (e.g., DoubleClick) collect large amounts of personal information from their users when browsing the web. The large scale collection and analysis of personal information constitutes the core business of most of these companies, which use it for commercial purposes. Our personal information can be used, for example, for precise targeting of ads \cite{ScroogledGMailScanWeb, AdStackEmailOptimizationPlatformWeb}, price discrimination \cite{mikians2012detecting, CreditCardsOffersWeb}, assessing our health and mental condition \cite{InsuranceDataPersonalFinanceWeb, InsurersTestDataProfilesWeb}, or assessing financial credibility \cite{FacebookFriendsCouldChangeYourCreditScoreWeb, ABCNewsGMASomeCreditCardCompaniesWeb, OutrageAsCreditagencyFacebookWeb}. Apart from that, the data can be accessed by government agencies and identity thieves. Some affiliate programs (e.g., pay-per-sale \cite{PayPerSaleAffiliateProgramsAffiliateDirectory}) require tracking to follow the user from the website where the advertisement is placed to the website where the actual purchase is made \cite{schmucker2011web}.

Personal information in the web can be voluntarily given by the user (e.g., by filling web forms) or it can be collected indirectly without their knowledge through the analysis of the IP headers, HTTP requests, queries in search engines, or even by using JavaScript and Flash programs embedded in web pages. Among the collected data, we can find information of technical nature, such as the browser in use, the operating system, the IP address, or even details about the underlying hardware, but also much more sensitive information, such as the geographical location of the users, their preferences or even the history of the web pages they visit. Unfortunately, the collected data do not stop here. For example, webmail services are known for scanning and processing user's e-mails, even if they are received from a user who did not allow any kind of message inspection.

This article surveys the literature on the methods currently used to track the user online by web services as well as their purposes, implications, and possible user's defenses. A significant majority of reviewed articles and accessed web resources are from years 2012 -- 2014. The selected material covers online blogs, original research papers, as well as small surveys about some particular tracking methods. The topic is treated by us holistically and systematized according to defined criteria and hierarchy, which aim at providing easy understanding and comparison of the approaches to track the users' online activity. A high-level non-technical overview of the most common tracking techniques is presented in Chapter 7 of \cite{tarnay2013research}.

Our survey can be useful for a broad range of readers. At first, it is going to increase the awareness of Internet users and make them able to both better protect their private data and to understand which data is collected from them and how the obtained information is being used. At second, it is, according to our best knowledge, the first work that surveys this topic holistically in a comprehensive way. Therefore, it can be used as a help for students and other researchers working on this topic. At third, we expect this survey to start a discussion about the privacy issues between the online providers, advertisers, Internet users, and regulatory agencies. The discussion should lead to developing some regulations (either external or self-regulations) and to throw out from the market unfair businesses that use offensive tracking or use the collected data for malicious purposes. That would strengthen both the Internet users and those companies that comply with acceptable tracking practices. Otherwise, if no regulations are being developed, the current online advertisement model is going to collapse. Namely, when users become aware that they are tracked in probably offensive ways, they will presumably start to use ad blocking tools, destroying the web economy and causing severe loss to themselves (e.g., by missing a potentially interesting offer directed to them or the necessity to pay for reading a news portal due to insufficient income from ads), to the ad providers, and to the services that use ads as the main income source.

This survey is structured in three main parts. At first, we present the known tracking mechanisms (Sections~\ref{sec:tracking_mechanisms_session_only}~--~\ref{sec:tracking_mechanisms_sophisticated}), afterwards we show their purposes and implications (Section~\ref{sec:implications_of_tracking}), and at the end, we discuss the possible defenses (Section~\ref{sec:general_tracking_auditing_and_defense_techniques}).

\begin{figure}[p]
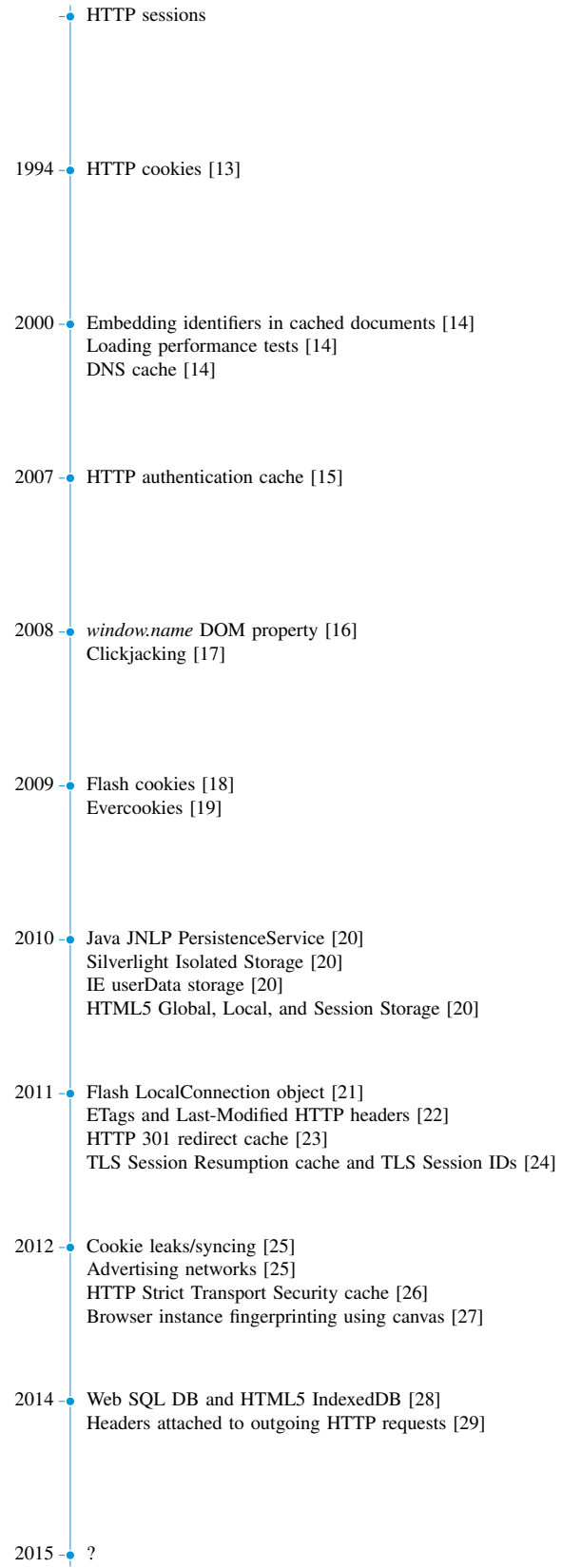

\begin{center}
\resizebox{!}{.9\textheight}{%
    \begin{vtimeline}[description={text width=10cm}, row sep=15ex, timeline color=cyan!80!blue]
    \mbox{} & HTTP sessions \endlr
    1994 & HTTP cookies \cite{GivingWebMemoryCostItsUsersPrivacyWeb} \endlr
    2000 & Embedding identifiers in cached documents \cite{felten2000timing} \newline
	Loading performance tests \cite{felten2000timing} \newline
	DNS cache \cite{felten2000timing} \endlr
    2007 & HTTP authentication cache \cite{TrackingUsersWithBasicAuthWeb} \endlr
    2008 & \emph{window.name} DOM property \cite{JavascriptSessionVariablesWithoutCookiesWeb} \newline
	Clickjacking \cite{YouDontKnowClickjackWeb} \endlr
    2009 & Flash cookies \cite{YouDeletedYourCookiesThinkAgainWeb} \newline
	Evercookies \cite{soltani2010flash} \endlr
    2010 & Java JNLP PersistenceService \cite{evercookieWeb} \newline
	Silverlight Isolated Storage \cite{evercookieWeb} \newline
	IE userData storage \cite{evercookieWeb} \newline
	HTML5 Global, Local, and Session Storage \cite{evercookieWeb} \endlr
    2011 & Flash LocalConnection object \cite{LocalConnectionActionScriptWeb} \newline
	ETags and Last-Modified HTTP headers \cite{PersistentUnblockableCookiesUsingHTTPHeadersWeb} \newline
	HTTP 301 redirect cache \cite{TrackingHTTPElieBurszteinWeb} \newline
	TLS Session Resumption cache and TLS Session IDs \cite{DisableTLSSessionResumptionTorWeb} \endlr
    2012 & Cookie leaks/syncing \cite{Roesner2012DDA22282982228315} \newline
	Advertising networks \cite{Roesner2012DDA22282982228315} \newline
	HTTP Strict Transport Security cache \cite{HSTSPersistenceAndPrivacyWeb} \newline
	Browser instance fingerprinting using canvas \cite{mowery2012pixel} \endlr
    2014 & Web SQL DB and HTML5 IndexedDB \cite{acar2014web} \newline
	Headers attached to outgoing HTTP requests \cite{HowVerizonsAdvertisingHeaderWorksWeb} \endlr
    2015 & ? \endlr
    \end{vtimeline}
}%
 \caption{Tracking mechanisms timeline based on their first documented occurrence}
 \label{fig:tracking_timeline}
\end{center}
\end{figure}

Sections~\ref{sec:tracking_mechanisms_session_only}~--~\ref{sec:tracking_mechanisms_sophisticated} present 19 main groups of methods used for user tracking. The timeline of the first reported occurrences of the particular tracking methods is shown in Figure~\ref{fig:tracking_timeline}. The technologies used to track the user by different tracking methods are shown in Table~\ref{tab:tracking_technologies}, while their tracking scope is summarized in Table~\ref{tab:tracking_scope}. Section~\ref{sec:tracking_mechanisms_session_only} describes the less intrusive techniques, which can track the user only during a single browsing session (e.g., session identifiers). Section~\ref{sec:tracking_mechanisms_client_storage} presents the most widely known generation of tracking techniques, which are based on a persistent storage on a user's computer; cookies fall into this category. They can be easily blocked by a wide range of available tools. However, as we show later, these mechanisms were also exploited to invent more invasive techniques, such as cookie leaks or syncing, and the aggregation of trackers into advertising networks. The explicit web-form authentication, with and without the support from cookies, is yet another technique widely used over the Internet. Nevertheless, it does not pose major privacy concerns, because it is accepted (or even chosen) by the user who must sign up to a service. There are, however, many intrusive techniques, which were designed and are intentionally used just to overcome the privacy protection mechanisms incorporated in the browsers. Such techniques include the use of plugin storages and standard HTML5 storages. Tracking methods exploiting the browser's cache are shown in Section~\ref{sec:tracking_mechanisms_cached_based}. The newly emerged fingerprinting techniques, which are able to identify the user without relying on any client-based storage and regardless of the private browsing mode enabled are shown in Section~\ref{sec:fingerprinting}. The last group of tracking methods presented in Section~\ref{sec:tracking_mechanisms_sophisticated} include the ones that are less expected by the user and that raise numerous ethical concerns. The tracking mechanisms are described together with the methods, which can be used for discovering and mitigating the particular threats.


\begin{table*}[t]
\caption{Technologies used to track the users}
\label{tab:tracking_technologies}
\centering
\tiny
\resizebox{2\columnwidth}{!}{%
\begin{tabular}{|r|l|l|}
\hline
\textbf{Section} & \textbf{Tracking mechanisms} & \textbf{Technologies} \\
\hline
\hline
\textbf{\ref{sec:tracking_mechanisms_session_only}}\mbox{~~~~} & \textbf{Session-only} &\\
\hline
\textit{A}\mbox{~~} & \textit{Session identifiers stored in hidden fields} & Web-server session \\
\hline
\textit{B}\mbox{~~} & \textit{Explicit web-form authentication} & Web-server session \\
\hline
\textit{C}\mbox{~~} & \textit{window.name DOM property} & HTML5, JavaScript \\
\hline
\hline
\textbf{\ref{sec:tracking_mechanisms_client_storage}}\mbox{~~~~} & \textbf{Storage-based} &\\
\hline
\textit{A}\mbox{~~} & \textit{HTTP cookies} & HTTP headers, JavaScript \\
\hline
\textit{B}\mbox{~~} & \textit{Flash cookies and Java JNLP PersistenceService} & Flash / Java \\
\hline
\textit{C}\mbox{~~} & \textit{Flash LocalConnection object} & Flash \\
\hline
\textit{D}\mbox{~~} & \textit{Silverlight Isolated Storage} & Silverlight \\
\hline
\textit{E}\mbox{~~} & \textit{HTML5 Global, Local, and Session Storage} & HTML5, JavaScript \\
\hline
\textit{F}\mbox{~~} & \textit{Web SQL Database and HTML5 IndexedDB} & HTML5, JavaScript \\
\hline
\textit{G}\mbox{~~} & \textit{Internet Explorer userData storage} & JavaScript \\
\hline
\hline
\textbf{\ref{sec:tracking_mechanisms_cached_based}}\mbox{~~~~} & \textbf{Cache-based} &\\
\hline
\textbf{\textit{A}}\mbox{~~} & \textbf{\textit{Web cache} }&  \\
\hline
1 & Embedding identifiers in cached documents & HTML5, JavaScript \\
\hline
2 & Loading performance tests & Server-side measurements, JavaScript\\
\hline
3 & ETags and Last-Modified headers & HTTP headers \\
\hline
\textbf{\textit{B}}\mbox{~~} & \textbf{\textit{DNS lookups}} & JavaScript \\
\hline
\textbf{\textit{C}}\mbox{~~} & \textbf{\textit{Operational caches}} &  \\
\hline
1 & HTTP 301 redirect cache & HTTP headers \\
\hline
2 & HTTP authentication cache & HTTP headers, JavaScript \\
\hline
3 & HTTP Strict Transport Security cache & HTTP headers, JavaScript \\
\hline
4 & TLS Session Resumption cache and TLS Session IDs & Web-server session \\
\hline
\hline
\textbf{\ref{sec:fingerprinting}}\mbox{~~~~} & \textbf{Fingerprinting} &\\
\hline
\textit{A}\mbox{~~} & \textit{Network and location fingerprinting} & IP address, server-based geolocation techniques, HTTP headers, HTML5, JavaScript, Flash, Java \\
\hline
\textit{B}\mbox{~~} & \textit{Device fingerprinting} & IP address, TCP headers, HTTP headers, JavaScript, Flash \\
\hline
\textit{C}\mbox{~~} & \textit{Operating System instance fingerprinting} & JavaScript, Flash, Java, ActiveX \\
\hline
\textit{D}\mbox{~~} & \textit{Browser version fingerprinting} & HTML5, JavaScript, CSS \\
\hline
\textit{E}\mbox{~~} & \textit{Browser instance fingerprinting using canvas} & HTML5, JavaScript \\
\hline
\textit{F}\mbox{~~} & \textit{Browser instance fingerprinting using web browsing history} & Server-side measurements, HTTP headers, JavaScript \\
\hline
\textit{G}\mbox{~~} & \textit{Other browser instance fingerprinting methods} & HTTP headers, JavaScript, Flash \\
\hline
\hline
\textbf{\ref{sec:tracking_mechanisms_sophisticated}}\mbox{~~~~} & \textbf{Other tracking mechanisms} &\\
\hline
\textit{A}\mbox{~~} & \textit{Headers attached to outgoing HTTP requests} & HTTP headers \\
\hline
\textit{B}\mbox{~~} & \textit{Using telephone metadata} & Smartphone malware \\
\hline
\textit{C}\mbox{~~} & \textit{Timing attacks} & HTML5, JavaScript, CSS \\
\hline
\textit{D}\mbox{~~} & \textit{Using unconscious collaboration of the user} & HTML5, JavaScript, CSS, Flash \\
\hline
\textit{E}\mbox{~~} & \textit{Clickjacking} & HTML5, JavaScript, CSS \\
\hline
\textit{F}\mbox{~~} & \textit{Evercookies (supercookies)} & Web-server session, HTTP headers, HTML5, JavaScript, Flash, Silverlight, Java \\
\hline
\end{tabular}
}%
\end{table*}


\begin{table*}[p]
\caption{Known tracking scope}
\centering
\resizebox{2\columnwidth}{!}{
\tiny
\begin{tabular}{|r|l|m{5cm}|}
\hline
\textbf{Section} & \textbf{Tracking mechanisms} & \textbf{Scope}\\ 
\hline
\hline
\textbf{\ref{sec:tracking_mechanisms_session_only}}\mbox{~~~~} & \textbf{Session-only} &\\
\hline
\textit{A}\mbox{~~} & \textit{Session identifiers stored in hidden fields} & Session id \\
\hline
\textit{B}\mbox{~~} & \textit{Explicit web-form authentication} & User id \\
\hline
\textit{C}\mbox{~~} & \textit{window.name DOM property} & Session id \\
\hline
\hline
\textbf{\ref{sec:tracking_mechanisms_client_storage}}\mbox{~~~~} & \textbf{Storage-based} &\\
\hline
\textit{A}\mbox{~~} & \textit{HTTP cookies} & Browser instance id \\
\hline
\textit{B}\mbox{~~} & \textit{Flash cookies and Java JNLP PersistenceService} & Operating system instance id \\
\hline
\textit{C}\mbox{~~} & \textit{Flash LocalConnection object} & Operating system instance id \\
\hline
\textit{D}\mbox{~~} & \textit{Silverlight Isolated Storage} & Browser instance id \\
\hline
\textit{E}\mbox{~~} & \textit{HTML5 Global, Local, and Session Storage} & Browser instance id \\
\hline
\textit{F}\mbox{~~} & \textit{Web SQL Database and HTML5 IndexedDB} & Browser instance id \\
\hline
\textit{G}\mbox{~~} & \textit{Internet Explorer userData storage} & Browser instance id \\
\hline
\hline
\textbf{\ref{sec:tracking_mechanisms_cached_based}}\mbox{~~~~} & \textbf{Cache-based} &\\
\hline
\textbf{\textit{A}}\mbox{~~} & \textbf{\textit{Web cache} }&  \\
\hline
1 & Embedding identifiers in cached documents & Browser instance id, browsing history \\
\hline
2 & Loading performance tests & Browsing history\\
\hline
3 & ETags and Last-Modified headers & Browser instance id \\
\hline
\textbf{\textit{B}}\mbox{~~} & \textbf{\textit{DNS lookups}} & Browsing history \\
\hline
\textbf{\textit{C}}\mbox{~~} & \textbf{\textit{Operational caches}} &  \\
\hline
1 & HTTP 301 redirect cache & Browser instance id \\
\hline
2 & HTTP authentication cache & Browser instance id \\
\hline
3 & HTTP Strict Transport Security cache & Browser instance id \\
\hline
4 & TLS Session Resumption cache and TLS Session IDs & Browser instance id \\
\hline
\hline
\textbf{\ref{sec:fingerprinting}}\mbox{~~~~} & \textbf{Fingerprinting} &\\
\hline
\textit{A}\mbox{~~} & \textit{Network and location fingerprinting} & IP address, user's country, city, and neighborhood \\
\hline
\textit{B}\mbox{~~} & \textit{Device fingerprinting} & Device id, IP address (entire or a part), operating system, screen resolution, timezone, list of system fonts, web browser, information about hardware (mouse, keyboard, accelerometer, multitouch capability, microphone, camera), TCP timestamps \\
\hline
\textit{C}\mbox{~~} & \textit{Operating System instance fingerprinting} & Operating system instance id, operating system version and architecture, system language, user-specific language, local timezone, local date and time, list of system fonts, color depth, screen dimensions, audio capabilities, access to the user's camera, microphone, and hard disk, printing support, hard disk identifiers, TCP/IP parameters, computer name, Internet Explorer product id, Windows Digital Product Id, installed system drivers, operating system instance id stored by a Java privileged applet \\
\hline
\textit{D}\mbox{~~} & \textit{Browser version fingerprinting} & Detailed browser version \\
\hline
\textit{E}\mbox{~~} & \textit{Browser instance fingerprinting using canvas} & Browser instance id \\
\hline
\textit{F}\mbox{~~} & \textit{Browser instance fingerprinting using web browsing history} & Browser instance id, browsing history \\
\hline
\textit{G}\mbox{~~} & \textit{Other browser instance fingerprinting methods} & Browser instance id, detailed browser version, supported formats of images and media files, preferred and accepted languages, list of browser plugins, browser user's language, browser dimensions, Flash version, screen resolution, color depth, timezone, system fonts, IP address, accepted HTTP headers, cookies enabled, supercookies limitations \\
\hline
\hline
\textbf{\ref{sec:tracking_mechanisms_sophisticated}}\mbox{~~~~} & \textbf{Other tracking mechanisms} &\\
\hline
\textit{A}\mbox{~~} & \textit{Headers attached to outgoing HTTP requests} & Customer's id\\
\hline
\textit{B}\mbox{~~} & \textit{Using telephone metadata} & Health (including mental) condition, religious believes, and addictions of a specific real person \\
\hline
\textit{C}\mbox{~~} & \textit{Timing attacks} & Boolean values dependent on the look of the website (e.g., if the user is logged in to a particular service), stealing any graphics embedded or rendered on the screen \\
\hline
\textit{D}\mbox{~~} & \textit{Using unconscious collaboration of the user} & Browsing history, browser instance id, user's location \\
\hline
\textit{E}\mbox{~~} & \textit{Clickjacking} & User's email and other private data, Paypal credentials, spying on a user by a webcam \\
\hline
\textit{F}\mbox{~~} & \textit{Evercookies (supercookies)} & Operating system instance id, browser instance id \\
\hline
\end{tabular}
\label{tab:tracking_scope}
}
\end{table*}

Section~\ref{sec:identification_of_the_tracked_user} shows how the users can be identified on the web and associated with their real names, e-mail addresses, phone numbers, or even street addresses.

Section~\ref{sec:implications_of_tracking} shows why tracking is being used and what the possible implications of it are for the users, e.g., by describing previously observed examples of price discrimination, assessing financial credibility, determining insurance coverage, government surveillance, and identity theft. We also show that in-website tracking can be used for non-invasive purposes as web analytics and usability tests, which are made in order to improve the user browsing experience.

Section~\ref{sec:general_tracking_auditing_and_defense_techniques} discusses the defense techniques and tools that can be used against more than one tracking method. The most popular are blocking the advertisement services, hiding the IP address by VPNs, the use of proxies or TOR, and using the private browsing mode. In addition to these, there exist more sophisticated methods, e.g., modification of the data sent over the network and execution blocking. The user can also choose to use privacy-focused search engines, anonymous mailing services and mail aliases, and can decide to frequently clear the browser cache and history. Apart from describing the methods and tools used for keeping the personal data away from being tracked, we also present here several tools that can be used for research purposes -- their main goal is to discover how and by what the user is being tracked on his desktop computer or smartphone, provide the information to the user, and visualize it in an accessible and easy to follow way.

Section~\ref{sec:the_future_of_tracking} presents the currently proposed future approaches to track the user and show that they can potentially increase the threats to the users' privacy. An example can be the solution being prepared by the Future of the Cookie Working Group from the Interactive Advertising Bureau (IAB), which relies on device-inferred, client generated, network-inserted, server-issued, and cloud-synchronized identifiers. Other examples of identifiers designed to have broad coverage are Google AdID and Microsoft Device Identifier.

Section~\ref{sec:conclusion} concludes the survey and summarizes the main findings.

\section{Session-only tracking mechanisms}
\label{sec:tracking_mechanisms_session_only}

In this section, we start to introduce several groups of tracking techniques. The historically first known tracking mechanisms relied on sessions. As we see in Table~\ref{tab:tracking_technologies} and Table~\ref{tab:tracking_scope}, the methods are quite simple and do not pose significant threats to the users.

\subsection{Session identifiers stored in hidden fields}

Before cookies became available in 1994 \cite{GivingWebMemoryCostItsUsersPrivacyWeb}, the only way to track a user except using the IP address was to pass his identifier to another website in the URL (GET method) or as a value in a hidden field of a web form (POST method) \cite{mckinley2008cleaning}. The identifier can be any string that is able to uniquely track the user during a single browsing session. It can be composed, for example, from a timestamp and a random number. The values submitted by a web form using POST, contrary to the values appended to the URL using GET, are neither persisted on the disk nor preserved in the browser history \cite{schmucker2011web}. Although this technique still works, its usefulness is limited to a single browsing session. However, the identifier can be passed to third parties \cite{HTTPAccessControlMozillaWeb} when no client-side programing language (e.g., JavaScript) is involved into the submission of the web form.

\subsection{Explicit web-form authentication}
\label{subsec:tracking_mechanisms_session_only_explicit_webform_authentication}

Another possibility to identify the user is to ask or require him to register on the website. Then, the resources provided by the website (e.g., webmail) are available only to the user who logs in. That make the user identification very easy and accurate. This method is independent of the web browser, operating system, or computer used, as well as of the place where the user is connected to the Internet. There are, however, two important issues connected to this method. At first, the user must authenticate every time he uses the website, which can be seen as burdensome to him. At second, the authentication is valid only within the current session after the user logs in. At third, it is impossible to track the user transparently -- the user is always aware that he is logged in and that everything he does can be recorded.

\subsection{\emph{window.name} Document Object Model (DOM) property}

The W3C Document Object Model (DOM) \cite{W3CDocumentObjectModelWeb} is a cross-platform interface for accessing and interacting with the content, structure, and style of web documents (i.e., HTML, XHTML, and XML). It is language-independent and it organizes all the objects in a tree structure. Each object (e.g., a \emph{window}) has a number of properties (e.g., a \emph{name}).

The \emph{window.name} DOM property \cite{HTTPCookieWikipediaWeb} can hold up to 2\,MB of data as a single string value. By using a JSON \emph{stringifier} \cite{JSONStringifyRevisitedWeb}, it is possible to pack multiple variables into one string, which can be subsequently stored on the user's computer. The \emph{window.name} property is resistant to page reloads and it is accessible from other domains as well, which gives the third-party content an opportunity to exchange the information with the first-party or with another third-party content. An implementation of cookieless session variables using this DOM property is shown in \cite{JavascriptSessionVariablesWithoutCookiesWeb}.

\section{Storage-based tracking mechanisms}
\label{sec:tracking_mechanisms_client_storage}

The next group of tracking mechanisms depend on explicit storing data on the users' computers. These methods seem to be the most commonly used ones. Generally, they are much more advanced than the session-based methods (see Table~\ref{tab:tracking_technologies}) and their abilities are also higher, as they are able to recognize the particular instance of a browser or an operating system, depending on the mechanism (see Table~\ref{tab:tracking_scope}). Each of these mechanisms presented the biggest threat to users' privacy directly after it was invented. Over time, browsers started to implement clearing of these storages on the user's request.

\subsection{HTTP cookies}

The most currently known method to identify a user is by using cookies -- small pieces of data (each limited to 4\,KB) placed in a browser storage by the web server \cite{RFC6265HTTPStateManagementMechanismWeb}. When a user visits a website for the first time, a cookie file with a unique user identifier (could be randomly generated) is stored on the user's computer. Then, the website can retrieve this identifier each time the user visits it, unless the user delete the cookie from his computer. This method of identification is very fast, does not require any kind of interaction from the user, and is completely transparent to him as the browser does not show any notifications when the cookies are set or read. However, its accuracy is limited to the situations when the user allows using cookies, does not clean the cookies cache in the browser, and always uses the same web browser to visit the particular website. There are two basic kinds of cookies: session cookies and persistent cookies. Session cookies expire when the user closes the web browser, while persistent cookies expire after a specified amount of time \cite{trackingAndIdentifyingTechReport}.

Cookies are set on the user's computer in 2 ways: either by a JavaScript using an API call, or by HTTP responses containing the \emph{Set-Cookie} header. Cookies can be read by the services in 2 ways as well. At first, they are automatically attached to the HTTP requests made to the domain to which the cookies belong using \emph{Cookie} headers. At second, they can be explicitly requested by a JavaScript API and then sent to the server in any way \cite{Roesner2012DDA22282982228315}.

Usually, tracking and non-tracking cookies can be distinguished based on their expiration time and length of the value field. Experiments described in \cite{li2014trackadvisor} show that more than 90\,\% of tracking cookies have their lifetime greater than 1 day, while such lifetimes concern only 20\,\% of non-tracking cookies. As the value contained by a cookie must be long enough to be able to distinguish each user, 80\,\% of tracking cookies have their values longer than 35 characters, while that concerns only 20\,\% of non-tracking cookies \cite{li2014trackadvisor}. In order to avoid this detection, websites can try to split the user's identifier in multiple cookies, but such behavior can be detected by taking into account the sum of the value lengths from all the cookies from the same third-party website \cite{li2014trackadvisor}. The authors of \cite{li2014trackadvisor} showed that based only on these attributes, their tracking recognition tool is characterized by precision of 99.4\,\% and recall of 100.0\,\%.

Some browsers or browser add-ons (for the list of defense techniques look at Section~\ref{sec:general_tracking_auditing_and_defense_techniques}) try to increase the users' privacy level by blocking third parties from setting and reading cookies (usually only from setting, however, Firefox also forbids reading). This limitation can be easily bypassed by redirecting the user by a JavaScript to the third-party website that will set or read the cookies; from this website, the user is redirected back to the one he originally visited. That way, the third-party content appears as coming from first parties. For this purpose, instead of redirecting the user, the website can use popup windows in which the third-party content is shown as coming from the first party \cite{blackhat2012ImplementingWebTrackingWeb}. As only Firefox is known from forbidding reading cookies by third parties by default, in other cases, it is enough just to set the cookies by a popup or a redirection in the first-party context, while they can be read directly by the third-party tracker \cite{Roesner2012DDA22282982228315}.

Research from 2011 shows that only around 30\,\% of users delete their cookies (both first party and third party) within a month from their acquisition \cite{TheImpactOfCookieDeletionWhitepaper}.

HTTP cookies can be used as a tracking mechanism alone or combined with other techniques.

\subsubsection{Explicit web-form authentication and cookies}

This method combines the web-form authentication (Section~\ref{subsec:tracking_mechanisms_session_only_explicit_webform_authentication}) and cookies in order to make the authentication more friendly to the user. The user registers on the website and logs in to be able to use the service as in the previous method. The user credentials are, however, stored in cookies on the user's computer. Thanks to that, the user does not need to log in again to the service as long as he allows using cookies, does not clean the cookies cache in the browser, and uses the same web browser. The authentication performed by the services owned by Google is a good example of this method.

\subsubsection{Cookie leaks / syncing}

During the experiment described in \cite{Roesner2012DDA22282982228315}, the authors observed a large number of cookie leaks. It means that cookies from one domain were passed to another domain, for example, as parameters of a request. This method is used, for example, by Microsoft to track the users on its domains (e.g.,  \emph{bing.com}, \emph{microsoft.com}, \emph{msn.com}, \emph{live.com}, and \emph{xbox.com}) and other websites which include resources from these domains. The cookies between these domains are exchanged regardless if the user visits both sites during one browser session or completely separately \cite{Roesner2012DDA22282982228315, TrackingTrackersMicrosoftAdvertisingWeb}. Cookie syncing, used for example by Google \cite{CookieMatchingGoogleDevelopersWeb} to better facilitate targeting and real-time bidding, enables third parties to exchange information about a user. This is possible by making requests containing the user identifier from one domain to another.

\subsubsection{Advertising networks}

Some websites embed a limited number of trackers, which act as aggregators for tracking services. For example, a very frequent tracker \emph{admeld.com} is known from making requests to other trackers, e.g., \emph{turn.com} and \emph{invitemedia.com}. The requests contain the visited website and the identifier assigned to that user by the aggregator \cite{Roesner2012DDA22282982228315}.

\subsection{Flash cookies and Java JNLP PersistenceService}

Local Shared Objects (LOSes) are used by Adobe Flash to store data on users' computers \cite{ayenson2011flash}. They can store 100\,KB of data by default, which makes their use for tracking more proper than using HTTP cookies, which can store only 4\,KB of data. Moreover, it is harder for the user to erase them than HTTP cookies, and they are accessible from all the browsers installed in the system, as all the instances of Adobe Flash plugins share the same storage directory. That makes them able to track the users across different browsers \cite{mittal2010user}. Apart from that, Flash cookies do not expire by default. They are stored in a disk directory as \emph{.sol} files.

Apart of the Local Shared Objects, Flash also supports Remote Shared Objects (RSOs). The cooperating Flash objects from the same domain can access the contents of locally persisted objects stored in a disk directory as \emph{.sor} files \cite{blackhat2012ImplementingWebTrackingWeb}.

Since version 10.3, Adobe Flash supports \emph{ClearSiteData} API, which removes the shared Flash objects. This API call is currently invoked by all main browsers while clearing the usual HTTP cookies. Additionally, the new Flash manager allows to block storing any Flash objects on the local computer \cite{verleg2014cache}.

Another option to store data is to use Java JNLP PersistenceService \cite{javaxJnlpWeb}, which facilitates storing data locally on the client system, also for applications that are running in the untrusted execution environment.

\subsection{Flash LocalConnection object}

The LocalConnection object supported by Flash can be used to communicate between different SWF files running on the same computer at the same time. Therefore, it also can be used to communicate and exchange values between Flash instances running in the normal and private browsing windows \cite{LocalConnectionActionScriptWeb}. This approach can be also combined with Flash cookies to pass the values from the cookies accessible in a normal browsing window to a Flash instance running in a private browsing window.

\subsection{Silverlight Isolated Storage}

This storage allows to save 100\,KB of information per site in the user's profile (as Flash cookies). However, this storage is disabled in private mode. The last version of Silverlight was announced by Microsoft to expire in 2021 \cite{verleg2014cache}. This storage can be cleaned only manually (by deleting files from a hidden folder in the file system or by using storage options in the Silverlight application). The Silverlight application also offers the possibility to disable the Isolated Storage \cite{verleg2014cache}.

\subsection{HTML5 Global, Local, and Session Storage}

HTML5 Global Storage \cite{DOMStorageGuideWeb} introduced by an early HTML5 draft offered some possibilities to store data by websites, but it was not implemented by any main browser due to the violation of the same-origin policy \cite{verleg2014cache}.

HTML5 Local Storage \cite{WebStorageWebLocalstorage}, which obeys the same-origin policy, provides yet another possibility to track the user. Placing the objects (key-value pairs) in the storage does not require any plugin. The objects are stored permanently (there is no automatic expiration) -- they persist until they are removed by the website or by the user. Additionally, an object can be as big as 5\,MB, which gives a significant advantage over both HTTP and Flash cookies \cite{lawson2011introducing}. The content of the Local Storage can be shared between different browser windows \cite{verleg2014cache}. The Local Storage is automatically emptied at the time when the cookies are cleared. The research performed in \cite{ayenson2011flash} showed that 17 among the top QuantCast.com ranking websites in 2011 used HTML Local Storage to store 60 key/value pairs. In several cases, the values stored in HTML Local Storage matched the HTTP cookies: \emph{twitter.com}, \emph{foxnews.com}, \emph{nytimes.com}, and \emph{cnn.com}. The match was usually seen with third-party websites, as \emph{meebo.com} (now being a part of Google), \emph{kissanalytics.com}, or \emph{polldaddy.com}.

HTML5 Session Storage \cite{WebStorageWebSessiontorage} is very similar to the Local Storage: it preserves the same-origin policy and the stored objects can be as big as 5\,MB. However, the objects are available only to the current browser window and are deleted when the window is closed \cite{verleg2014cache}.

\subsection{Web SQL Database and HTML5 IndexedDB}

Web SQL Database \cite{WebSQLDatabaseWeb} uses SQLite instead of the local file system to store data on the client side. Although it was implemented by all main browsers, its development was stopped in favor of HTML5 IndexedDB \cite{IndexedDatabaseAPIWeb}. HTML5 database storages operate under the same conditions as the local storage, thus, the privacy impact is the same.

The first report of using HTML5 IndexedDB to rebuild Flash and HTTP cookies is from 2014 \cite{acar2014web}. The authors found that a script from \emph{weibo.com} stores an item in this database, which directly matches the content of the Flash cookie \emph{coosimg.sinajs.cn/stonecc\_suppercookie.sol} and the corresponding HTTP cookies from \emph{weibo.com} and \emph{sina.com.cn}.

\subsection{Internet Explorer userData storage}

Internet Explorer userData storage \cite{userDataBehaviorWeb} is a proprietary technique introduced in Internet Explorer 5.5 and declared obsolete in Internet Explorer 7, however, it still works in Internet Explorer 11. It allows to store up to 64\,KB of data in the XML format \cite{verleg2014cache}.

\section{Cache-based tracking mechanisms}
\label{sec:tracking_mechanisms_cached_based}

Another group of tracking methods also use client-based storage. But in contrast to the previous group that used storages explicitly designed for preserving data, this group exploits possibilities to identify browser instances and determine the previously visited websites (see Table~\ref{tab:tracking_scope}) by the use of various caches.

\subsection{Web cache}
\label{subsec:tracking_mechanisms_cached_based_web_cache}

Before 2010, the browsing history could be easily obtained using DOM API in an automatic way. The attacker could set the color used by the browser to display visited and non-visited links according to his wish. Then, JavaScript could be used to generate links pointing to a specific destination (e.g., \emph{http://www.cnn.com}), which the attacker would like to check in the browsing history. The color of the generated links could be read by JavaScript and compared with the color previously set to be used by both the visited and non-visited links. Using this technique, an attacker could test 10\,000 to 30\,000 links for presence in the browser history \cite{janc2010web, wondracek2010practical}. This threat, however, was fixed in all popular browsers after a solution was proposed in \cite{baron2010preventing}. The fix relies on returning the style of all hyperlinks by API calls as non-visited regardless if the links are visited or not.

Despite fixing this threat, web browsing history still can be obtained by a number of ways. As it is shown in the following paragraphs, the most known automatic techniques use browser caching. When a browser downloads an object (e.g., an image), it is usually stored in the browser cache for faster display when the user visits the website again. Therefore, when a user browses a website, the website can easily determine if this user visited it before (so that this object is pulled from the cache) or not (so that the object is downloaded from the server). When an advertiser has his objects on many websites, he can easily compare them with the cached copies and determine which pages were visited \cite{juels2006cache, ayenson2011flash}. Exploiting the web cache can be done in several ways.

\subsubsection{Embedding identifiers in cached documents}

The first possibility to identify a user is to make the user request an HTML file, which contains the embedded identifier. The identifier can be stored in an invisible \emph{div} and subsequently read from the browser cache using the \emph{div id} property \cite{blackhat2012ImplementingWebTrackingWeb}. As the cached files can be included by any website, the values stored in them can be used across multiple services \cite{felten2000timing}.

\subsubsection{Loading performance tests}

Websites can use JavaScript to detect the time of loading any object (e.g., an image) from any URL. The loading time can be measured by JavaScript and reported to the host service, which can evaluate if the object is present in the browser cache or not. That way, by testing on an object that is always accessed when a user uses the particular website (e.g., company logo), the script can assess if the website was previously accessed \cite{felten2000timing}. In case it is not possible (e.g., JavaScript is blocked), the offensive service can try to use another technique: load at first a file from itself, then a file from the website whose presence in the browser cache is tested, and then again a file from itself. The time used to load the file from the tested website can be calculated by subtracting the times between loading the files from the offensive service itself \cite{felten2000timing}.

\subsubsection{ETags and Last-Modified HTTP headers}

To enable user identification using the web cache, entity tags (ETags) \cite{fielding1999hypertext} or \emph{Last-Modified} HTTP headers \cite{fielding1999hypertext} can be used. The HTTP header provided with the first download of an object contains the following fields: \emph{Last-Modified}, \emph{ETag}, \emph{Cache-Control}, and \emph{Expires}. The \emph{ETag} field can be as long as 81864 bits and, therefore, it is sufficient to store the user identifier, which became used by the tracking companies. The \emph{Last-Modified} header \cite{fielding1999hypertext} is shown to accept any random string, not only a valid date \cite{PersistentUnblockableCookiesUsingHTTPHeadersWeb}, which also started to be used for tracking purposes. When a user visits the website again, the browser sends the \emph{If-Modified-Since} and \emph{If-None-Match} headers as a part of the HTTP request (these headers contain values stored in the \emph{Last-Modified} and \emph{ETag} fields from the previously cached document, respectively). The web server then checks if the cached copy is not outdated \cite{juels2006cache}. If the cached copy is still valid, the server returns a very short response with the \emph{HTTP 304 Not Modified} status. In case if it is outdated, a new document is returned.

The first use of ETags to track the users was observed in 2011, when hulu.com employed KISSmetrics service to rebuild HTTP and HTML5 cookies. To avoid tracking by ETags, the user must clear the browser cache before each visit on the website that compares the particular ETag. Tracking is also possible during a single private browsing session, as the cache is kept until the last browser window is closed \cite{ayenson2011flash}. The method using the \emph{Last-Modified} headers does not have any problem with bypassing web proxies, which can be problematic for ETags \cite{fielding1999hypertext}.

\subsection{DNS cache}

Yet another automatic method use the possibility of JavaScript to indirectly cause a DNS lookup and measure its time. In case the website was previously accessed, the corresponding entry exists in the DNS cache, which significantly reduces the lookup time \cite{felten2000timing}.

\subsection{Operational caches}

Operational caches are components used to store information related to operations made by the web browsers, rather than to store the copies of the downloaded elements. Such stored information includes permanent redirects, authentication credentials, or a list of domains that must be used together with the HTTP Strict Transport Security (HSTS).

\subsubsection{HTTP 301 redirect cache}

The \emph{HTTP 301 redirect} mechanism was designed to tell the browser that the particular queried resource is permanently available at another URL. The browser caches the redirect and uses it instead of the original URL during the following attempts to visit the website. That could be used if an existing page was moved somewhere else or if the initial link was a shortened URL, generated to conserve space or provide better visibility. However, this mechanism can be used to track the user by third parties. The first time the user visits a service (which can be observed by the service by the format of the URL used to access it), the service generates a \emph{HTTP 301 redirect} to the same URL as initially used, but with the attached identifier for the particular user. As this request is cached, the browser is going to use this URL (together with the attached user identifier) during the next attempts to access the service \cite{TrackingHTTPElieBurszteinWeb}. The redirection can be made in an iframe, so it is transparent to the users \cite{verleg2014cache}.

\subsubsection{HTTP authentication cache}

There are two commonly used HTTP authentication mechanisms: \emph{Basic access authentication} and \emph{Digest access authentication} \cite{schmucker2011web}. When a user enters his credentials on a website using one of these mechanisms, the website stores the credentials temporarily, so they can be automatically submitted in the HTTP \emph{authorization header} to the server upon next requests, so the session can be identified. Jeremiah Grossman on his website \cite{TrackingUsersWithBasicAuthWeb} presents several methods that use JavaScript to force the browser to authenticate to the web server in a way that no popup with an authentication request is presented to the user. When the browser is authenticated and it receives \emph{401 (Unauthorized)} status code, it will cache the credentials and send them with each subsequent HTTP request.

\subsubsection{HTTP Strict Transport Security cache}

HTTP Strict Transport Security (HSTS) \cite{jackson2012http} is yet another mechanism which can be used to create a cookie-like storage. Mikhail Davidov on his blog \cite{HSTSPersistenceAndPrivacyWeb} described a possible implementation of such a storage. The current standard of HSTS specifies that the server can ask over a secure connection to make all the future connections to its domain using HTTPS instead of HTTP. The server indicates when the HSTS mode will expire and whether the sub-domains should be covered by the policy. Wildcard SSL certificates can be issued for a server to cover multiple subdomains within the same top-level domain. When a user visits websites in the HSTS mode, for each of the subdomains, a single entry is made in the HSTS database on the user's computer. The website is able to encode the user's identifier character-by-character in different subdomain names, for example, an identifier ``\emph{DBAS}'' can be encoded into \emph{1-D.addomain.com}, \emph{2-B.addomain.com}, \emph{3-A.addomain.com}, and \emph{4-S.addomain.com}. When some resources are requested from these domains (e.g., a pixel image) using HTTPS while HSTS is on for the domain, these subdomains are stored into the HSTS database. Then, during the later user's visit on the website, a JavaScript code can use a brute-force method to query iteratively by HTTP the resources from all the possible combinations of the subdomains, e.g., \emph{1-A.addomain.com}, \emph{1-B.addomain.com}, \emph{1-C.addomain.com}, and \emph{1-D.addomain.com}; in this case, during the first 3 tries, the URL will not get translated into HTTPS, as no entry in the HSTS database exist. However, when asking for the resource from \emph{1-D.addomain.com}, HTTP is going to be rewritten by HTTPS, which indicates that the first character of the user's identifier is \emph{D}. Mikhail also claims that Google Chrome clears the HSTS database at the same time as the cookies are deleted, while Firefox 4 stores the HSTS database in a separate place, which is not easy to clear.

\subsubsection{TLS Session Resumption cache and TLS Session IDs}

TLS session resumption cache is used to store TLS/SSL session IDs \cite{dierks2008transport}, which are sent by the server to the client during the \emph{hello} message, in order to use them when connecting to the host the next time. That avoids full TLS handshakes, which reduces latencies and the CPU usage \cite{TLSSessionResumptionImplementationsWeb}. It is shown in \cite{DisableTLSSessionResumptionTorWeb} that both the TLS session resumption cache and TLS/SSL session IDs can be used for third-party tracking and, thus, their implementation was fixed in Tor Browser.

\section{Fingerprinting}
\label{sec:fingerprinting}

Fingerprinting is a group of methods using a broad range of technologies (see Table~\ref{tab:tracking_technologies}), which is able to discover the widest spectrum of properties from all the methods (see Table~\ref{tab:tracking_scope}).

A fingerprint (a unique identifier of a device, operating system, browser version or instance) can consist of one or more values which can be read by the web service when the user browse its website. That way, the user can be tracked on multiple different websites belonging to different entities, which is not directly possible to be done using cookies. No cookies are generated and the user does not need to log in, so the tracking is transparent to the user and it works regardless if the browser accepts cookies or not. Therefore, an average user does not have any means to know if he is tracked and the user does not know how to prevent that (although, it is possible in some extent by disabling the support of JavaScript, Java, and Flash, however, it does not prevent passive fingerprinting). A study from 2010 made by Yahoo \cite{HowManyUsersHaveJavaScriptDisabledWeb} reveals that only around 1\,\% of users disabled the support of JavaScript in their browsers. Some of the properties used for fingerprinting can be discovered passively from the network traffic (IP address, operating system, user agent, language, HTTP accept headers), while others require (or can be also obtained) by active approaches (operating system, CPU type, user agent, timezone, clock skew, display settings, installed fonts, installed plugins, enabled plugins, supported MIME types, cookies enabled, third-party cookies enabled) \cite{mayer2012third}. There are numerous websites which show us extensive amount of information that can be automatically collected from the user's computer \cite{MyBrowserInfoWeb, SystemDetailsWeb, PanopticlickWeb, BrowserSPYWeb}. According to \cite{nikiforakis2013cookieless}, 40 out of top 10\,000 Alexa websites, including \emph{skype.com}, use fingerprinting scripts from BlueCava \cite{BlueCavaWeb}, Iovation \cite{IovationWeb}, or ThreatMetrix \cite{ThreatMetrixWeb}. The most popular website categories that use fingerprinting were found to be porn (15\,\%) and dating (12.5\,\%). The next sections are going to present different categories of fingerprinting methods, define their scope, coverage, and potential defenses.

\subsection{Network and location fingerprinting}

One of the easiest features to be determined based on the headers of incoming HTTP requests are the global network address and the IP-based geographical location of the user. By using network tools, the service is able to identify the name of the domain and the user's Internet Service Provider. The presence of a proxy server can be detected by asking about that explicitly in an HTTP request \cite{HTTPHeaderFieldDefinitionsWeb}. Proxies set in the browser can be bypassed by Flash applets to discover the real IP address of the client \cite{nikiforakis2013cookieless}, which is done by Iovation \cite{IovationWeb} and ThreatMetrix \cite{ThreatMetrixWeb}, two of the most popular fingerprinting companies. Therefore, anonymity-providing applications, e.g., Tor Browser, tend to block Flash. Several important features can be returned by JavaScript functions, as the internal IP address \cite{ipcalfWeb} and GPS coordinates \cite{HTML5GeolocationW3SchoolsWeb}. The connection performance (e.g., the download and upload speeds, round-trip time, and jitter) can be calculated either by JavaScript \cite{SpeedOfMeWeb} or by Flash \cite{SpeedtestOoklaWeb}. Finally, Java applets can be used to detect the presence of a firewall \cite{WhatAppletsCanAndCannotDoWeb}.

Traditionally, the user's location can be inferred from the IP address. There are many free databases (e.g., IP To Country Database \cite{IPtoCountryDatabaseWeb}), which facilitate discovering the country and city where the IP address should be placed. A comprehensive survey on the possible geolocation mechanisms that can be used based on the IP address is shown in \cite{wang2011towards}. In \cite{balakrishnan2009s}, it was shown that the IP-based location is not accurate for mobile networks. The authors claimed that over 90\,\% of mobile devices in Seattle have IP addresses located more than 600 miles from Seattle. Apart from that, VPN or Tor can be used to hide the real global IP address, as it is masked by the IP address of the VPN gateway or a Tor exit relay \cite{HowSocialMediaToppledADictatorWeb}. Therefore, geolocation is a research topic and there are many existing papers tackling with this problem by different methods. For example, a geo-inference method, based on the history and cache of visited websites, is shown in \cite{jia2014know}.

\subsection{Device fingerprinting}

A novel cross-browser device fingerprinting method based only on JavaScript and using passively collected information is shown in \cite{boda2012user}. The user identifier is derived from the first two octets of the IP address, version of the operating system, screen resolution, timezone, and the list of the standard fonts available in the system, which are universal for all the web browsers. Apart from on the user identifier, the authors collected for research purposes some other variables, as the \emph{User-Agent} string and the real IP address. The list of system universal fonts was obtained solely by JavaScript. Thanks to such a design, the authors achieved browser independence (the user identifier does not depend on any browser-specific feature), plugin independence, and cross-domain tracking (given that the user has its IP address assigned from the same pool). Furthermore, using the visual inspections, they were able to recognize returning users who used multiple browsers, changed their screen resolutions, timezones, or the IP addresses.

The Turkish Ministry of National Education \cite{TurkishMinistryNationalEducationWeb} was shown in \cite{acar2013fpdetective} to be the only known \emph{.gov} website, which uses extensive device fingerprinting. A Flash object embedded on the website extracts and sends back to the server the detailed information about the mouse, keyboard, accelerometer, multi-touch capability, microphone, camera, and system fonts.

A remote device fingerprinting method was described in \cite{kohno2005remote}. This method is based on the clock skew on fingerprinted machines and it uses TCP timestamps. Therefore, it is able to recognize the device regardless of the distance from the device, location of the device (e.g., cable or wireless network), or the presence or absence of NATs, which is an advantage over methods based on MAC or IP/ICMP. However, although the clock skews are constant during time, they are not enough unique to be able to identify a device with a high probability, as many different machines can have the same skew, especially, when they use NTP or other time synchronization methods. It is why this method can be used to deny the existence of the particular machine in a network at the certain time, but it cannot be used to confirm that the particular machine at this time was present in the network. The authors tried to de-anonymize a previously anonymized Caida trace with 11\,862 IP addresses. The identification accuracy based on the clock skew was 57.66\,\% or 87.65\,\% depending on the maximal TCP timestamp option clock skew differences given as a parameter.

\subsection{Operating System instance fingerprinting}

The version and architecture (32/64 bit) of the operating system can be identified both by JavaScript and Flash. JavaScript also facilitates recognizing the system language and system user-specific language, local timezone \cite{TimezoneDetectionWeb}, and the local date and time up to 1 millisecond \cite{JavaScriptDatesWeb}. The list of installed fonts can be detected both by JavaScript \cite{FontDetectByJavaScriptWeb} and Flash, which is performed by scripts and Flash objects of three popular large fingerprinting companies \cite{nikiforakis2013cookieless}: BlueCava \cite{BlueCavaWeb}, Iovation \cite{IovationWeb}, and ThreatMetrix \cite{ThreatMetrixWeb}. The color depth and screen dimensions can be detected both by JavaScript \cite{DetectScreenSizeWeb} and Flash \cite{SystemFlashClassWeb, CapabilitiesFlashClassWeb}. Flash can also detect whether the system has audio capabilities, whether the access to the user's camera and microphone has been prohibited or allowed, whether the system does or does not support printing, and whether read access to the user's hard disk has been prohibited or allowed \cite{SystemFlashClassWeb, CapabilitiesFlashClassWeb}. Standard Java applets can open, read, and save files on the client, thus, they can be used to store tracking information. Privileged Java applets can run outside the security sandbox and have extensive capabilities to access the client, obtaining all the information available to a standard system application \cite{WhatAppletsCanAndCannotDoWeb}. Another possibility used to read the values from the operating system is to use ActiveX controls. Research shown in \cite{nikiforakis2013cookieless} demonstrated that BlueCava \cite{BlueCavaWeb} and Iovation \cite{IovationWeb} use them to read the hard disk identifier, TCP/IP parameters, computer name, Internet Explorer product identifier, Windows installation date, Windows Digital Product Id, and the installed system drivers.

\subsection{Browser version fingerprinting}

The detection of the web browser by the self-reported \emph{User-Agent} field in HTTP headers is considered to not be reliable, as various browsers and other applications are able to obfuscate this string or claim to be a particular browser \cite{unger2013shpf, mulazzani2013fast, mowery2011fingerprinting}.

In \cite{unger2013shpf}, the authors introduced new browser version fingerprinting methods, called CSS and HTML5 fingerprinting. These methods are able to recognize the family and version of the web browser, however, they cannot be used to identify the particular browser instance. CSS fingerprinting uses the differences between the implemented CSS properties, CSS selectors, and CSS filters to distinguish one browser version from another. The overview of CSS supported properties by different web browsers can be found in \cite{CanIUseWeb}. The support of the particular property can be confirmed by JavaScript in two ways: by asking if the property exists or trying to set the particular property and then trying to read its value \cite{unger2013shpf}. HTML5 fingerprinting relies on the differences between how the web browsers implement the standard. In \cite{unger2013shpf}, 242 new tags, attributes, and features of HTML suitable for fingerprinting were detected. New HTML tags introduced in HTML5 amounted for 30 of them, while the rest consisted of new features and attributes for already existing tags.

The JavaScript engine was used for browser identification in \cite{mulazzani2013fast}. As the methods relying on CSS or HTML5, it is able to detect the family and version of the browser, but not the particular instance. ECMAScript \emph{test262} is a test suite built to check how the JavaScript implementations follow the ECMA-262 standard of the JavaScript language. It consists of 11552 tests and it executes in around 10 minutes on a desktop PC or 45-60 minutes on a mobile device. The particular tests are passed or failed depending on the browser. The authors propose two methods for the identification of various web browsers: finding a minimal subset of \emph{test262} that is able to distinguish one web browser from another, or building a decision tree based on the results of the tests. The second approach allows to include more tests in the comparison while reducing the script execution time.

The performance of the JavaScript engine was used in \cite{mowery2011fingerprinting} to detect the web browser version, the operating system, and even the architecture of the computer. The authors constructed a suite of 39 JavaScript tests. Each test was separately executed 5 times and its execution time was recorded. A delay of 800\,ms was added between every launching of these script to minimize the effect of any cleanup processes made by the browser. The minimal execution times were taken into account while creating the browser profiles, so that the differences in the use of the system resources by other processes could not impact the results.

\subsection{Browser instance fingerprinting using canvas}

HTML5 introduced an area of the screen, which can be used to draw text or images programmatically. Currently, there are two defined graphics contexts: \emph{2d} and \emph{webgl}. Using the \emph{2d} context, the browser can use \emph{fillRect}, \emph{lineTo}, and \emph{arc} methods to draw basic figures or \emph{fillText} to draw the text. The text can be formatted using styling methods similar to the ones found in CSS \cite{mowery2012pixel}.

A special kind of fingerprinting is \emph{canvas fingerprinting}, which relies on using the browser canvas API to draw invisible graphics. There are several methods to retrieve the produced graphics. The \emph{getImageData} of the \emph{2d} context returns an object containing the RGBA values (as integers) for every pixel of the requested rectangular region of the canvas. The second method, \emph{ToDataURL}, returns a Base64 encrypted string encoding the entire content of the canvas in the requested format (e.g., PNG) \cite{mowery2012pixel}. As every browser instance draws the graphic differently, the Base64 encrypted string differs as well from one browser to another, which results in the possibility of extracting a unique hash from the string and using it as a browser identifier. The rendered image depends on the operating system, installed fonts, graphics card and its drivers, and the browser itself, due to font rasterization, anti-aliasing, smoothing, API implementations, and the physical display \cite{mowery2012pixel}. According to \cite{acar2014web}, canvas fingerprinting is the most commonly used fingerprinting method, present on more than 5.5\,\% of top 100\,000 Alexa websites. The authors found out that the examined JavaScripts used for fingerprinting the following methods: \emph{fillText} and \emph{strokeText} to write the text, and \emph{ToDataURL} to read the image data. There are also some other methods which could be used for fingerprinting (but not directly): \emph{MozFetchAsStream}, \emph{getImageData}, and \emph{ExtractData}, however, they were not investigated in the paper. A JavaScript was considered as fingerprinting in case if both methods (writing the image and reading the image data) came from the same script (URL), if the image contained more than one color and was bigger than 16x16 pixels, and if the image was not requested in a lossy compression format, which could remove the details. In case more images were drawn by the script, the dimension criteria concerned the sum of the dimensions of all the images. The rendered image can be accessed only by the same origin as drew it, otherwise, a \emph{SecurityError} exception is returned.

fingerprintJS \cite{FingerprintjsValveGitHubPages}, a free JavaScript library, also supports canvas fingerprinting and the returned value is the same in the normal and private browsing modes \cite{ValveFingerprintjsGitHub}.

The fingerprinting scripts from \emph{addthis.com} were responsible for the majority (95\,\%) of the fingerprinting attempts on the analyzed websites \cite{acar2014web}. This script used the perfect pangram ``\emph{Cwm fjordbank glyphs vext quiz}'' as the drawn text string. The text was printed two times, once, using a particular font, second time, using the default browser font (the script requests an non-existing font). Additionally, the script checks for drawing Unicode support (it tries to print the \emph{U+1F603} smiling face character). It also checks for canvas \emph{globalCompositeOperation} support (it sets or returns how a new image are drawn onto an existing one) and it draws two rectangles to check if a specific point is in the path by the \emph{isPointInPath} method.

The same font (e.g., Arial) looks differently at different operating systems. To make the website use the same font regardless of the hardware and software used by the computer, the website can use WebFonts instead. A CSS3 rule called \emph{@font-face} allows us to load any web font. We can also use a WebFont Loader \cite{WebFontLoaderWeb} library designed to load web fonts using only JavaScript. The web fonts can also be used to draw on the browser canvas.

Normally, web browsers use the GPU for rendering 3D or even 2D graphics. In \cite{mowery2012pixel}, the authors proved that machines with the same browser, operating system, and graphics card, always render the image in the same way and the produced fingerprint is identical. However, various combinations of them render the image differently, which can be used to identify the browser, OS, and graphics card on the fingerprinted computer.

Tor Browser is the only known browser that protects the user from canvas fingerprinting by asking the user every time a canvas function which can be used to read image data is invoked. If not, an empty string is returned \cite{TorAntiFingerPrintWeb}. In addition to that, Tor Browser takes actions against using the Web Graphics Library (WebGL) to obtain a fingerprint \cite{TorBrowserDraftWeb}. At first, WebGL can reveal the underlying driver and optimizations. At second, this library can be used for performance fingerprinting. This document also mentions keystroke fingerprinting as a kind of a biometric fingerprinting method, which relies on measuring the key strike and flight times.

A possible defense scheme for fingerprinting would be to use hardware acceleration to produce the pixels visible on the screen for performance reasons, but to use software rendering libraries such as Pango to return platform-independent strings whenever the site wants to read the produced image \cite{mowery2012pixel}.

\subsection{Browser instance fingerprinting using web browsing history}

Methods for obtaining web history were described in Section~\ref{subsec:tracking_mechanisms_cached_based_web_cache} and Section~\ref{subsec:tracking_mechanisms_sophisticated_using_unconscious_collaboration_of_the_user}.

Research performed on 368\,284 Internet users \cite{olejnik2012johnny} revealed that it is possible to fingerprint 42\,\% of them by testing which websites out of 50 pre-defined ones were visited. In case if the test considered a set of 500 websites, it was possible to fingerprint 70\,\% of the users. The websites included in the test were taken out of 500 top Alexa ranked websites and from the top 4\,000 QuantCast popular websites. The authors also converted the history profiles of visited websites by each user to category profiles. Around 88\,\% of the profiles appeared to be unique. It was shown that the history profiles are stable during the time. Even after clearing the browser data, the new profiles were in 38\,\% of cases strongly correlated to the old ones.

Another approach to identify and track users based on the history of visited websites is shown in \cite{wondracek2010practical}. All the major social networks allow their members to be organized in groups: public or closed ones. Every group is associated with a numerical identifier, which is a part of the URL pointing to the user's activity in the group, as browsing the posts. This URL is saved in the browsing history. The authors found that the groups to which a user belongs can be used not only to fingerprint the particular user, but also to identify his real name. The built-in functionality of social networks can be used to crawl various groups and list their members. During the 23 days of the experiment, the authors managed to crawl over 43.2 million group members from 31\,853 Facebook groups. The same experiment was repeated on Xing; 1.8 million users from over 6\,500 groups were crawled. Based on that, it is possible to create a database of the users and the associated groups. When a malicious script launches a web history attack and discovers which groups were visited, it is possible to correlate the outcomes with the list of groups to which the particular users belong. For the Xing network, 42.06\,\% of members had an unique fingerprint based on the groups to which they belong. For the rest of them, the fingerprint allows to narrow down the list of potential users. In the real-world experiment and launching a test website, the authors were able to de-anonymize 1\,207 (37.3\,\%) out of 3\,717 Xing users \cite{wondracek2010practical}.

\subsection{Other browser instance fingerprinting methods}

A lot of browser instance-specific information is known to be detectable by using HTTP requests and the analysis of the responses \cite{HTTPHeaderFieldDefinitionsWeb}: the detailed version of the web browser, supported formats of images and media files, or preferred and accepted languages. Apart from the web browser version, JavaScript can list the browser plugins \cite{PluginDetectJsWeb} and identify the browser user's language. Browser dimensions and Flash version can be recognized both by JavaScript \cite{DetectScreenSizeWeb, SWFObjectWeb} and Flash \cite{SystemFlashClassWeb, CapabilitiesFlashClassWeb}. fingerprintJS \cite{FingerprintjsValveGitHubPages} is a free JavaScript library which is able to fingerprint the web browser based on its agent string, screen resolution, color depth, installed plugins with supported mime types, timezone offset, local storage, and session storage. Studies performed in \cite{eckersley2010unique} show that the installed plugins, screen resolution, timezone, system fonts, and user-agent strings yielded altogether an entropy of 18.1 bits. In \cite{yen2012host}, the authors claim that the \emph{User-Agent} string (which is supposed to be store the detailed information about the used browser version) alone has the entropy of 11.59 bits, while combined with the IP address results in 20.29 bits of entropy. It was also shown in \cite{eckersley2010unique} that if a user changes a fingerprinted property (as updates the plugin or changes the screen resolution), his current fingerprint can be matched to the earlier one by a heuristic algorithm with the accuracy of around 99\,\%.

In \cite{nikiforakis2013cookieless}, the authors evaluated mechanisms used to identify the browser's instance by 3 popular large fingerprinting companies: BlueCava \cite{BlueCavaWeb}, Iovation \cite{IovationWeb}, and ThreatMetrix \cite{ThreatMetrixWeb}. They discovered that all the companies use both JavaScript and Flash for fingerprinting.

\subsubsection{Panopticlick project}

The browser fingerprinting project Panopticlick \cite{PanopticlickWeb} uses different features to create and check the fingerprint of the particular instance of a web browser. After we got our browser fingerprinted by Panopticlick, the fingerprint appeared to be unique among 4\,561\,261 browsers tested so far. A quite surprising is the fact that the user agent ``\emph{Mozilla/5.0 (X11; Linux i686) AppleWebKit/537.36 (KHTML, like Gecko) Chrome/35.0.1916.153 Safari/537.36}'' was common for one of 10\,834.35 browsers tested so far, the set of installed browser plugins was common for one of 5\,715.87 browsers, while the set of installed system fonts was common for one of 4\,561\,261 tested browsers. Other information taken into account while computing the browser fingerprint (accepted HTTP headers, timezone, color depth, screen size, cookies enabled, and supercookies limitations) were common for 1--31 tested web browsers. Summarizing, our browser was assessed to have a fingerprint that conveys at least 22.12 bits of identifying information. As shown, these easy to obtain properties are sufficient to accurately identify the concrete instance of the web browser \cite{APrimerOnInformationTheoryAndPrivacyWeb, eckersley2010unique}.

\section{Other tracking mechanisms}
\label{sec:tracking_mechanisms_sophisticated}

The last group contains different methods, which use various methodologies (see Table~\ref{tab:tracking_technologies}) to discover various properties (see Table~\ref{tab:tracking_scope}).

\subsection{Headers attached to outgoing HTTP requests}

Normally, when a web browser follows a link, the \emph{Referer} field in the HTTP header contains the URL of the currently browsed website. If the URL contains any appended data using the GET technique (e.g., search terms, login), the information is also passed in the HTTP request.

Verizon Wireless is known from attaching a special header \emph{X-UIDH} to every outgoing HTTP request \cite{HowVerizonsAdvertisingHeaderWorksWeb}. This header is static for a long time; the average changing time was reported to be around one week. Thanks to the header send from all the mobile devices using Verizon's wireless (including mobile) networks, all the websites are able to track the users even if they use very enhanced privacy settings. The web services then use Verizon's API to buy information about the subscribers' demographic and geographic segments.

\subsection{Using telephone metadata}

According to the studies shown in \cite{MetaPhoneTheSensitivityWeb}, the telephone metadata (e.g., call logs possessed by default by governments of most countries) is highly sensitive. The authors developed an Android application, which was capturing phone metadata. Using the call logs, the researchers were able to determine the health (including mental) condition of the person, the religious believes, and addictions. As many users place their phone numbers in publicly accessible directories (e.g., Google+, Facebook), the scientists could associate around 18\,\% of the private numbers with the real users' identities. Summarizing, mobile applications, being granted permissions to access the call logs and other sensitive data, are a serious threat to the users' privacy.

\subsection{Timing attacks}

Differences in time taken to render different DOM trees can be used to determine boolean values, e.g., if the user has an account on a tested website \cite{kotcher2013cross}. The authors found that WebKit renders the website by traversing the entire tree of \emph{RenderElements} from the root to leaves in order to resolve the pixel depth and it paints the element from the furthest away to the nearest. The users' login status (or any other state) can be determined by checking the color of particular pixels. The color of the pixel can be checked based on the time taken for the shader to output the final value \cite{kotcher2013cross}. The same paper proposes a pixel-stealing attack using CSS, which was shown to have high accuracy. This kinds of attacks, which can be used to read the browser history or even steal any graphics embedded on the website, are broadly described in \cite{stone2013pixel}.

\subsection{Using unconscious collaboration of the user}
\label{subsec:tracking_mechanisms_sophisticated_using_unconscious_collaboration_of_the_user}

Apart from the automatic solutions, the browsing history can be detected in cooperation with the user, who is unaware of that fact. Four easily implementable methods were proposed in \cite{weinberg2011still}. The ideas rely on producing images pretending to be CAPTCHAs, which the user need to type into a field below the image and submit to the service before some facilities or actions on the website are possible. The images are composed of words, letters, or other font-based symbols, which are in fact links to the destinations we are going to check if they were previously visited by the user. The color of the particular symbol is set to be the same as the background in case if the link was visited; otherwise, the color is set to be distinct than the background, so that the symbol can be easy recognized by the user. The produced CAPTCHA image is usually covered by a transparent image to make it more difficult to the user to recognize that this CAPTCHA is composed of hyperlinks. The first of the proposed techniques use word CAPTCHA; each word is a link to one tested website. The second solution uses LCD-like characters. Every character is composed of 4 different symbols with a nearly transparent gray background covering each other. Each of these symbols (except one, which is always black) is drawn black if the corresponding link is visited or white if non-visited. That way, depending on the combination of visited and non-visited links, the symbols compose a different character. The third solution use a gaming approach. A drawn chessboard grid is filled with pawns, which are in fact images rendered by text or SVG shapes (so their color can be controlled by CSS), which color depends on that if the corresponding link was visited or not. The user is asked to click on the visible pawns. As the website does not know on how many pawns the user needs to click, a time-out must be set after which the user is able to proceed further. The fourth presented method requires the user to select two puzzle images from which the shown resulting image can be composed. The resulting composite image is created using four SVG shapes, whose color depends on if the user visited the associated hyperlinks.

Additionally to these four universal and easy to implement solutions, a method based on a web camera is shown. Contrary to the previous ones, this method is limited to be used on websites, where the people normally uses their web cameras. The background color of the website depends on the condition if the user previously visited the tested website or not. The light of the background is reflected by the user and captured by the web camera. This method is limited to day-time estimations (as the paper proved, the accuracy is low in case if it is too dark in the room) and testing one particular website at a time. Although the background can change to test various websites, it cannot change too frequently to annoy the user \cite{weinberg2011still}.

There are also other techniques, which can be used to track the users and require their intervention in order to work. Such techniques use HTML5 protocol and content handlers, and W3C Geolocation API \cite{GeolocationAPISpecificationWeb}.

\subsection{Clickjacking}

Clickjacking is a method to present a sensitive website element out of context, so the user acts out of context. Several approaches to clickjacking, which can lead to compromising user's anonymity, stealing user e-mails and private data, and spying on a user by webcam are shown in \cite{huang2012clickjacking}. These techniques were used to conduct Tweetbomb \cite{ClickjackingTweetbombWeb} and Like-jacking \cite{ClickjackingWikipediaWeb} attacks. The clickjacking attacks were divided into 3 categories based on what is compromised: target display, pointer, or temporal integrity. The first method (compromising target display integrity) usually uses either hiding the target element by setting the opacity to 0 \cite{ClickjackingSecTheoryWeb} or covering the target by an unclickable opaque decoy \cite{ClickjackingWithPointerEventsWeb}, so that the user clicks the attackers link. Alternatively, a partial overlay can be made \cite{ClickjackingDetailsWeb, sirdarckcatWeb} or the target can be cropped \cite{sirdarckcatWeb}. The second method (compromising pointer integrity) uses the CSS \emph{cursor} property to hide the default cursor and draw a fake cursor somewhere else \cite{CursorjackingAgainWeb} or use a custom cursor icon \cite{ProofOfConceptCursorJackingWeb} to trick the user that the cursor is in another position. The third method (compromising temporal integrity) use the fact that people require a few hundred milliseconds to react \cite{RaceConditionsInSecurityDialogsWeb, OnDesigningUIsForNonrobotsWeb} in order to manipulate the target, e.g., move it after the users hovers the cursor over a button in order to click it \cite{ClickjackingAttacksUnresolvedWeb, XPInstallMozillaBugWeb}.

There exist several approaches to protect the user from clickjacking \cite{huang2012clickjacking}. Target elements (e.g., Facebook \emph{Like} buttons) started to require the user to confirm its decision \cite{FacebookAddsLikejackersWeb}. Randomizing the user interface layout \cite{hill2012adaptive}, e.g., Paypal \emph{Pay} button, makes it more difficult to discover where it is located in order to cover it by another element. Finally, the target element can disallow to be embedded in an \emph{iframe} by using JavaScript \cite{rydstedt2010busting} or \emph{X-Frame-Options} \cite{IE8ClickJackingDefensesWeb}.

\subsection{Evercookies (supercookies)}
\label{subsec:tracking_mechanisms_sophisticated_evercookies}

In \cite{soltani2010flash}, it was shown that in 2009, 54 out of the 100 most popular Quantcast websites used Flash cookies to rebuild removed HTTP cookies. The authors found that 41 of the Flash cookies had their content identical to the corresponding HTTP cookies. In \cite{ayenson2011flash}, it was found that ETags and the HTML5 Local Storage are also used to rebuild deleted HTTP cookies. In \cite{acar2014web}, the authors found that 10 out of 200 the most popular Alexa websites used Flash cookies to rebuild HTTP cookies. Flash cookies are shared between different browsers by using Adobe Flash plugin and, therefore, they can be used to expand the current user's identity and rebuild HTTP cookies even in a newly installed web browser. They also detected the cases where an HTTP cookie rebuilt by one domain is passed to another domain. The \emph{evercookie} project \cite{evercookieWeb} shows how to construct a resilient JavaScript evercookie (also called a supercookie), which is using various storages in order to survive, rebuild after deletion, or even reproduce in other browsers used at the same computer. The following storages are used by this special evercookie: HTTP Cookies, Local Shared Objects (Flash Cookies) \cite{LocalSharedObjectWikipediaWeb}, Silverlight Isolated Storage \cite{IsolatedStorageWeb}, web history \cite{CSSHistoryKnockerWeb}, ETags \cite{fielding1999hypertext}, web cache, \emph{window.name} DOM property \cite{HTTPCookieWikipediaWeb}, Internet Explorer userData storage \cite{userDataBehaviorWeb}, HTML5 Session Storage \cite{WebStorageWebSessiontorage}, HTML5 Local Storage \cite{WebStorageWebLocalstorage}, HTML5 Global Storage \cite{DOMStorageGuideWeb}, HTML5 Database Storage via SQLite \cite{WebSQLDatabaseWeb}, HTML5 IndexedDB \cite{IndexedDatabaseAPIWeb}, Java JNLP PersistenceService \cite{javaxJnlpWeb}, Java CVE-2013-0422 exploit (applet sandbox escaping) \cite{CVE20130422Web}, and finally, storing cookies in RGB values of auto-generated, force-cached PNGs using HTML5 Canvas tag to read pixels (cookies) back out.

The authors also considered introducing support for caching in the HTTP Authentication \cite{franks1999http}, using Java to produce a unique key based on the NIC info, and Google Gears \cite{mckinley2008cleaning}, which was discontinued by Google in 2011. Although, in these cases, a user intervention is required for the method to work.

Tor Browser Design and Implementation Draft \cite{TorBrowserDraftWeb} reveals various places, which can be potentially used for storing the cookies and, therefore, the data is erased while the user switches to another identity: searchbox and findbox text, HTTP auth, SSL state, OCSP state, site-specific content preferences (including HSTS state), content and image cache, offline cache, cookies, DOM storage, DOM local storage, the safe browsing key, and the Google WiFi geolocation token.

An article from 2011 \cite{FlashCookiesPrivacyAshkanSoltaniWeb} claimed that KISSmetrics \cite{KISSmetricsCustomerIntelligenceWebAnalyticsWeb}, a third-party service, is using HTTP cookies, Flash cookies, ETags, the userData storage, and the HTML5 local storage to create a persistent evercookie being able to recreate the missing parts if deleted. However, the official response from the company to the published article denied these facts \cite{OfficialKISSmetricsResponseDataCollectionPracticesWeb}. The company declares that it does not track users across different websites and the only mechanism used for tracking are first-party HTTP cookies. The use of ETags or any other persistent tracking object is denied. Moreover, the company is shown to respect the Do Not Track header, by avoiding tracking any information about the user, even during a single browsing session.

\section{Identification of the tracked user}
\label{sec:identification_of_the_tracked_user}

Being able to detect and track the digital identity of a user is valuable for many reasons. But even more profitable could be the ability to associate the digital identity of a user with the real identity (name, surname, social security number, etc). As shown in \cite{sweeney2001computational}, it is possible to unambiguously identify 87\,\% of the USA population based on only 3 attributes: the date of birth, gender, and ZIP code. That can cause a severe security threat to the users. If a user decides to reveal his true identity to one service (e.g., during the registration of an e-mail account), he risks being recognized by all the other entities with which the database of user data is shared based on the formally anonymous data, as the date of birth and gender. Five ways of identifying the user and associating him with his online activities, which are the base for our classification, were proposed in \cite{ThereIsNoSuchThingAsAnonymousOnlineTrackingWeb}.

\subsection{Legitimate first-party services tracking as third parties}

The tracking third party can be also a first party. Such a situation exists when we are browsing websites while we are logged into services as Facebook or Google. The Facebook \emph{Like} or Google \emph{+1} buttons embedded on various websites report the visit to Facebook or Google, respectively.

\subsection{Leaking information to third parties}

The identifier can leak from first parties to third parties by a number of ways. The identifier or the e-mail address can be provided in the \emph{Referer} field in HTTP headers. The \emph{Request-URI} header can also contain sensitive information as gender, ZIP code, or interests. Usernames or real names can be embedded in page titles or contained in shared cookies resulting from hidden third-party servers. The information collected by all these different means can be combined by a third party. For example, if a request to the third party contains a user identifier or an e-mail address in the \emph{Referer} field in HTTP headers and attached third-party cookie, the third party can easily associate the cookie with the particular identifier or e-mail address. In \cite{mayer2012third}, the authors observed several forms of leaking the identifying information to third parties as a parameter in a URL: Ads contained on the Home Depot website sent the first name and e-mail address of the user to 13 companies; when a user entered a wrong password on the Wall Street Journal Website, his e-mail address was sent to 7 companies; finally, video-sharing website Metacafe sent user's first name, last name, birthday, e-mail address, physical address, and phone numbers to 2 companies when the user changed his user settings.

In \cite{krishnamurthy2011privacy}, the results from examining over 100 popular non-social networking websites were shown. Fully 56\,\% of them directly leaked personal information to third parties. When considering leaking the user identifier, the number of leaking websites grew to 75\,\%, as the user identifier was passed by 48\,\% of the websites. The e-mail address was leaked by 13\,\% of them. The research was based on inspecting the HTTP requests, responses, and POST requests against predefined pieces of data entered previously to the user profiles, e.g., e-mail addresses, names, and zip codes. Most of the data leaked by the \emph{Referer} field in HTTP headers and by passing automatically cookies to third-party web servers operating under the domain of the first party.

\subsection{Selling information to third parties}

The first party can sell the identity to a third party. For example, an advertising data provider Datalogix buys data from consumer marketing companies and data compilers \cite{PrivacyDatalogixWeb}. These data include names, postal addresses, email addresses, and demographic and behavioral information, such as past purchases. The data are used to create profiled audiences and measure the efficiency of advertising. Datalogix also claims to receive and collect information about consumers’ visits to websites, such as the browser types, IP addresses, types of ads served, and dates of ad deliveries \cite{PrivacyDatalogixWeb}.

\subsection{Using web hacks}

A third-party content on the first-party website can use hacks in order to get the sensitive information. In 2010, it was demonstrated how to obtain the real name of the people associated with a particular Google account using Google Docs (this vulnerability was quickly fixed after being announced \cite{HowGoogleDocsLeaksYourIdentityWeb}). As a document edited online using Google Docs displays the names of the people who have the document opened, it was sufficient to create a fake website, which contained an invisible iframe (dimensions 0x0) embedding a publicly available Google document. If the victim was logged into his Google account, he could be identified by a script running on the fake website and looking who was recently attached to the list of people viewing the document.

\subsection{Intended deanonymization}

In 2013, a professor from Harvard University was able to re-identify 42\,\% of anonymous people participating in the Genome Project using census data and public records \cite{sweeney2013identifying}. Scientists also identified 30.8\,\% of Twitter and Flickr users using data correlation between online accounts \cite{narayanan2009anonymizing}. A Stanford University student discovered that almost 50\,\% of the 185 biggest websites share usernames and other sensitive information with online advertisers by putting them into the URL, which is sent back to the advertisers. That includes YouTube and Twitter \cite{SexDrugsITWorldWeb}. The Stanford Security Lab conducted a larger scale research and found out that a popular dating website, OkCupid, leaks information to advertising networks Lotame and BlueKai about the drinking, smoking, drug habits, ethnicity, kids, pets, location, and more \cite{TrackingTheTrackersWeb}. The Center for Internet and Society at Stanford Law School discovered that Epic Marketplace, a member of the self-regulatory Network Advertising Initiative (NAI) is stealing the users' history by a JavaScript \cite{TrackingTrackersToCatchHistoryThiefWeb}. The list of checked URLs contained 15\,511 entries, including pages related to getting pregnant and fertility, menopause, repairing bad credit, and debt relief.

An approach to link multiple user accounts by using only their usernames is shown in \cite{perito2011unique}. The uniqueness of a username (the probability that it refers to the same user across different online services) was estimated based on its entropy. The model was extended to cover also matching two different usernames to the same person. The ground truth was obtained by Google Profiles, on which the users are able to link their other accounts on other services. A Markov-Chain classifier trained on over 10 million of usernames gathered from eBay and Google showed that the entropy of both distributions is higher than 35 bits. The authors also revealed how the users construct their usernames. Over 70\,\% of the usernames contained the first name and/or the last name of the user. Around 30\,\% of the usernames are result of the concatenation of the first and last names without adding any additional character.

KISSmetrics \cite{KISSmetricsCustomerIntelligenceWebAnalyticsWeb}, introduced by us in Section~\ref{subsec:tracking_mechanisms_sophisticated_evercookies}, is an analytic service, which claims to be able to link all the possessed data with real users, even if they use multiple devices and browsers. It advertises itself as being able to track all the activities done by the users. Examples of the information contained by KISSmetrics databases are: the real name, gender, age, social networks logins and IDs, which websites were browsed and when, which e-mails were read, shared content in social networks, and purchased items in web stores. However, as mentioned before in Section~\ref{subsec:tracking_mechanisms_sophisticated_evercookies}, KISSmetrics use only first-party cookies and, therefore, the tracking cannot span among multiple services.

\section{Purposes and implications of tracking}
\label{sec:implications_of_tracking}

Tracking is commonly used by various entities for a wide range of purposes. The obtained information is not always used directly by the tracker - a very common practice is that the collected data are sold to other parties (e.g., insurance companies or online stores) or accessed by government agencies and identity thieves. Next, we describe a few examples of the most frequently shown applications of web tracking.

\subsection{Online advertising}

Web tracking was initially developed in order to better facilitate marketing and increase sales profit. Within the time, it emerged into sophisticated techniques as behavioral tracking, audience segmentation, and targeting. For these purposes, tracking is used nowadays by majority of websites. In this section, we will show several examples of how tracking is implemented.

GMail uses words from the sent and received email messages to display targeted ads. The received e-mails are scanned despite that the sender did not give an explicit the permission to do that. GMail scans the content of the entire inbox to identify the themes and trends for ad targeting \cite{ScroogledGMailScanWeb}. According to the Google Terms of Service: ``\textit{Our automated systems analyze your content (including emails) to provide you personally relevant product features, such as customized search results, tailored advertising, and spam and malware detection. This analysis occurs as the content is sent, received, and when it is stored}''. \cite{GoogleTermsOfServiceWeb}.

AdStack \cite{AdStackEmailOptimizationPlatformWeb} facilitates including advertisements in marketing e-mails. When a user opens a specially prepared e-mail, the advertisement is downloaded in real-time based on the information known about the specific user.

Affiliate programs (e.g., pay-per-sale \cite{PayPerSaleAffiliateProgramsAffiliateDirectory}) require tracking to follow the user from the website where the advertisement is presented to the website where the actual purchase is made \cite{schmucker2011web}.

\subsection{Web analytics and usability tests}

Web tracking can also be done for some internal purposes of the website host. There are two common applications of the internal tracking of a user by a website: web analytics and usability tests \cite{schmucker2011web}. These services, however, do not impose any threat to the user, as all the user's activity data are kept within a single website. The usability tests can be used to record and play back cursor movement paths, measuring the time spent by the user on the particular positions in a questionnaire, or other issues, which are important in order to improve the website in the future \cite{schmucker2011web}.

In \cite{jang2010empirical}, the authors performed a study of keyboard and mouse tracking on the Alexa top 1300 websites and they identified 115 websites, which were suspicious of user tracking. They found that 7 out of the 10 top identified websites (e.g., \emph{microsoft.com}) append additional event-handlers to track the clicks and send the information to a third-party server. A behavior tracking software from \emph{tynt.com} was used on 7 out of the 115 identified websites. The software, apart from its intended job of appending the URL of the website to the content copied by the user to the clipboard, sends the copied content to \emph{tynt.com}, which allows to profile the user according to the used information.

\subsection{Assessing financial credibility}

Many companies that assess creditworthiness use the online activity of the user as one of the criteria \cite{FacebookFriendsCouldChangeYourCreditScoreWeb}. For example, Lenddo determines if our Facebook friends are late with payment of their loans. If so, our credibility is automatically lowered proportionally to the degree of intimacy in the relation between the friends. Kreditech uses data from our Facebook, eBay, or Amazon accounts, and the user's location to assess if the user is credible or not. Kabbage requires the borrowers to grant it access to their PayPal, eBay, and other payment accounts. In addition to that, the creditworthiness can be significantly improved by linking the Facebook and Twitter accounts, which gives sense that the social network contacts are taken into account by the algorithm.

In 2009, Kevin Johnson reported to have his credit limit in American Express of 10\,800\,\$ lowered to 3\,800\,\$ after he shopped online in Walmart. American Express claimed that it was due to the fact that many other Walmart customers have problem with paying the credit back \cite{ABCNewsGMASomeCreditCardCompaniesWeb}.

Schufa, the largest German ranking company, wanted to use the data from Facebook, LinkedIn, and Twitter accounts to determine the connection between different people and to track changes of their locations \cite{OutrageAsCreditagencyFacebookWeb}.

\subsection{Price discrimination}

Tracking also can be used to modify the advertised price of products according to the estimated financial situation of the potential customers.

In \cite{mikians2012detecting}, it was proved that the displayed price differs based on the geographical location of the user visiting a website (up to 166\,\%), on the affluence of the user (up to 400\,\%), and on the referrer -- the website from which the user accessed the website selling the product (up to nearly 50\,\%). The authors collected data during 20 days, performing some actions on websites belonging to 200 online vendors. No discrimination was noticed based on the operating system or browser used by the customer.

An article \cite{CreditCardsOffersWeb} claims that the interest rates on credit cards offered by several companies vary depending on who is looking at the credit card offer on the web. For example, it was observed that the Chase Sapphire card was advertised with interest rates 13.24\,\% and 12.24\,\%. Capital One Financial Corporation uses the calculations made by a company called [x+1] to decide which credit cards should be showed to their visitors in the first place. [x+1] extensively uses various online tracking techniques to obtain and assemble information about the Internet users. [x+1] claims that it can access and analyze thousands of pieces of information about a single user, of which the most valuable are the ZIP code and the date of birth, as people living in the same region tend to have similar income and living habits \cite{CuttingEdgeWSJWeb}.

In 2010, Devin found out that Capital One Financial Corporation differentiates the interests for car loans based on the browser used by the prospective customer (3.5\,\% for Firefox, 2.7\,\% for Safari, 2.3\,\% for Chrome, and 3.1\,\% for Opera) \cite{CapitalOneMadeMeDifferentLoanOffersWeb}.

Orbitz Worldwide Inc. differently sorts out the hotel advertisements depending on the type of computer used by the customer. Orbitz found that Mac users tend to spend around 30\,\% more on hotel bookings than PC users. Using that fact, more expensive hotels are advertised to Mac users, while the cheaper ones to the PC users. Furthermore, the search results are sorted according to the estimated top amount the user is going to pay for the room \cite{OnOrbitzMacUsersWeb}.

Matt Ilardo was shown different prices when booking a car by Hotwire from his work (88\,\$) and home (117\,\$) computers. The prices were stable in time on both computers \cite{AreOnlineTravelAgenciesElliottBlogWeb}, which proves that the discrimination was made based on the user's identity associated with each browser, not on the time the booking was made.

\subsection{Determining the insurance coverage}

Our online activity tells a lot about our lifestyle, interests, habits, and hobbies. That information can be used by numerous companies and institutions to assess the risk of providing us an insurance.

Allfinanz and TCP LifeSystems develop software, which is able, through different marketing data, to assess what is the risk that a certain person gets cancer soon or becomes a victim of an accident. The data is sourced from product warranties, consumer surveys, magazine subscriptions, and credit-card spending. Based on the assumed lifestyle, some insurance companies modify the frequency of medical checks of their customers \cite{InsuranceDataPersonalFinanceWeb}.

Acxiom Corporation, one of the biggest data-collecting companies, is known from buying data from online publishers about the types of online articles read by subscribers. These data is correlated with publicly available information from social networks and other online profiles. Acxiom is presenting to have in average 1500 pieces of data about around 96\,\% of Americans and 700 millions of people globally \cite{InsurersTestDataProfilesWeb}.

\subsection{Impact on the job market}

It is estimated that around 90\,\% of employers in the US make background check of their employees (or job applicants). At the same time, the data found online is very often of bad quality, outdated, and confusing different persons having the same name \cite{APIMPACTWhenYourCriminalPastWeb}. Finland’s Data Protection Ombudsman banned employers from googling the job applicants after the case when an employer refused to hire a person on the ground that he participated in a mental health conference \cite{FinnishEmployersCannotGoogleApplicantsWeb}.

\subsection{Government surveillance}

The data obtained by user tracking is a valuable source of information for government agencies and law enforcement authorities. Between January and June 2014, the US government made 12\,539 requests for 21\,576 person's information from Google, including search history, and Google complied with 84\,\% of them \cite{UnitedStatesGoogleTransparencyReportWeb}. Google also actively scans the images that pass through GMail accounts to see if they match up with known child pornography \cite{TheVergePornWeb}.

Unique identifiers contained by tracking cookies make the government surveillance capabilities more powerful. The government agents, granted access to Internet links, are able to use cookies in order to distinguish flows originated by different users sharing the same Internet connection. The flow of the unique cookies exposes among others how the users change their locations during time. They also can be used to denounce the users who granted access to the network in a non-authorized way \cite{HowNSAPiggyBacksThirdPartyTrackersWeb}. According to the internal NSA presentations revealed by Edward Snowden in 2013 \cite{NSAUsesGoogleCookiesForHackingWeb}, the American NSA and British GCHQ use this technique to investigate the online activity of the users. Google \emph{PREFID} cookie was used for this purpose, as it is set by Google when a user visits its properties directly or indirectly (e.g., by an advertisement or a \emph{+1} button provided by Google services). The presentation also shows that Special Source Operations (SSO), a division of NSA, collaborates with private companies to collect data from their systems as well as from the Internet backbone. The logins, cookies, and Google \emph{PREFIDs} obtained by SSO were later shared with Tailored Access Operations, another division of NSA, which engages in offensive hacking operations, and with the British intelligence agency GCHQ \cite{NSAUsesGoogleCookiesForHackingWeb}. Another presentation revealed by Snowden showed that NSA uses Doubleclick cookies to identify TOR users. That technique was assessed to be more efficient than decrypting TOR tunnels, for which the government does not have sufficient computational power. On the other hand, many TOR users seem to not clean cookies when they switch their browser to communicate with TOR, so their identity can be easy exposed by matching the cookies sent during the TOR session with the cookies sent when the communication was not encrypted \cite{NSAUsesGoogleCookiesForHackingWeb}.

Another NSA program, HAPPYFOOT, was designed to map Internet addresses to their physical locations in a precise way. Mobile devices are able to determine their location in many ways: by using GPS, WiFi, or cellular towers. The location is sent to Google or other companies for tailored advertising. By capturing the Internet traffic, NSA gathers almost 5 billion records a day on the locations of cellphones around the world. That also allows NSA to track how particular people travel and gain the knowledge about their mutual relations by revealing co-travelers \cite{NSATrackingCellphoneLocationsWorldwideWeb}.

Citizenfour \cite{HomeCITIZENFOURWeb}, a documentary thriller based on the life of Edward Snowden and the NSA spying scandal, had its US and UK premieres in October, 2014. The name of the movie originated from a nickname of Snowden (\emph{Citizen Four}) used by him for the purpose of mailing with journalists before revealing the compromising information.

\subsection{Identity theft}

Finally, the collected information can be used to steal our identity.

Carnegie Mellon University studies showed that data revealed by people in the Internet, when combined together, are in many cases sufficient to predict the social security number of these people. That opens the doors to steal someone's identity, as the social security number is widely used to authenticate on sensitive websites (e.g., banking or loan services) \cite{FaceRecognitionStudyFAQWeb}.

The 2012 Identity Fraud Report revealed that LinkedIn, Google+, and Facebook users are much more likely to become victims of a fraud than other people. The reason could be that 68\,\% of them share their date of birth, 63\,\% share their high school name, 18\,\% their phone number, and 12\,\% the name of their pet \cite{IdentityFraudRoseReportWeb}. LinkedIn users or people who frequently check in using their GPS position were shown to be more than twice prone to be a fraud victim than the others \cite{UsingLinkedInBusinessInsiderWeb}.

\subsection{Prevalence of tracking}

Third-party tracking is a form of tracking performed by resources from other services that the one explicitly visited by the user. For example, when a user browses \emph{DR.dk}, his activity is tracked by Facebook by the mean of \emph{Like} buttons embedded on the sites belonging to \emph{DR.dk}. Third-party trackers (e.g., Doubleclick) are considered as a serious privacy threat, as they can collect and accumulate browsing statistics through many different websites. In \cite{li2014trackadvisor}, the authors found that around 46\,\% of home pages of the websites from the top 10\,000 Alexa ranking are monitored by at least one third-party tracker. In particular, one third of the requests sent to third-party websites was sent to a tracker. Google was responsible for tracking on 25\,\% of examined websites, Facebook on 13\,\%, and Twitter on 5\,\%. A newly released study \cite{falahrastegar2014anatomy} showed that the presence and dominance of local tracking third parties is different in different geographical regions. In Europe, East Asia, Oceania, and South America, the distribution of third parties is almost even, while in Turkey and Israel, the local third parties have a much stronger market position. German and Russian trackers are shown to be present among the top sites in every country. Websites located in the Middle East contain mostly European and American trackers. The highest number of third-party trackers was observed on websites from Qatar (814), Korea (769), and Hong Kong (726), while the lowest on the websites from the United Kingdom (397), Jordan (330), and Belgium (274), out of 6817 trackers instances noticed over 6497 examined pages in total.

A comprehensive study about the various characters of the third-party tracking domains is shown in \cite{krishnamurthy2009privacy}. The authors specify 4 different types of trackers. The first type uses only third-party cookies, e.g., \emph{doubleclick.net}. The second type uses first-party cookies together with a JavaScript code used to interrogate the cookies, which is included at the website. An example of such kind of a tracker is \emph{Google Analytics}. It was shown that nearly 60\,\% of all examined first-party servers used first-party cookies set by a third-party JavaScript. The third type of trackers combines both previous approaches and use both third-party cookies and first-party cookies together with a JavaScript code. An example can be \emph{quantserve.com}. The last, fourth type, is just used to serve advertisements, while the actual tracking process is performed by another service. An example is \emph{2mdn.net}, which is supplied with information by \emph{doubleclick.net}.

\section{Methods and tools for avoiding and auditing tracking}
\label{sec:general_tracking_auditing_and_defense_techniques}

There are several strategies, methods, and tools, which can be used to discover and defend against several different tracking techniques. The possible defense strategies against each of the known tracking techniques is shown in Table~\ref{tab:defense_strategies}. Many of these strategies make use of the existing tools (summarized in Table~\ref{tab:tracking_defense_tools}), while the rest require from the user taking individual actions (summarized in Table~\ref{tab:other_enhancement_techniques}). In this section, we compare and evaluate the possible solutions. Another useful list of methods that can be used against tracking is shown in \cite{TrackingTrackersSelfHelpToolsWeb}.


\begin{table*}[p]
\caption{Tracking defense strategies}
\centering
\resizebox{2\columnwidth}{!}{
\tiny
\begin{tabular}{|r|l|m{4.8cm}|}
\hline
\textbf{Section} & \textbf{Tracking mechanisms} & \textbf{Defense strategies}\\ 
\hline
\hline
\textbf{\ref{sec:tracking_mechanisms_session_only}}\mbox{~~~~} & \textbf{Session-only} &\\
\hline
\textit{A}\mbox{~~} & \textit{Session identifiers stored in hidden fields} & No defense known; although, not needed \\
\hline
\textit{B}\mbox{~~} & \textit{Explicit web-form authentication} & Do not authenticate \\
\hline
\textit{C}\mbox{~~} & \textit{window.name DOM property} & Block JavaScript execution \\
\hline
\hline
\textbf{\ref{sec:tracking_mechanisms_client_storage}}\mbox{~~~~} & \textbf{Storage-based} &\\
\hline
\textit{A}\mbox{~~} & \textit{HTTP cookies} & Disallow cookies (all, third-party, or selectively) \\
\hline
\textit{B}\mbox{~~} & \textit{Flash cookies and Java JNLP PersistenceService} & Disallow Flash cookies, block Flash execution / block Java applets \\
\hline
\textit{C}\mbox{~~} & \textit{Flash LocalConnection object} & Block Flash execution \\
\hline
\textit{D}\mbox{~~} & \textit{Silverlight Isolated Storage} & Disable Isolated Storage, block Silverlight execution \\
\hline
\textit{E}\mbox{~~} & \textit{HTML5 Global, Local, and Session Storage} & Block JavaScript execution, disallow cookies (HTML5 storages run under the same policy as cookies) \\
\hline
\textit{F}\mbox{~~} & \textit{Web SQL Database and HTML5 IndexedDB} & Block JavaScript execution, disallow cookies (these storages run under the same policy as cookies) \\
\hline
\textit{G}\mbox{~~} & \textit{Internet Explorer userData storage} & Block JavaScript execution, disable the userData storage in IE \\
\hline
\hline
\textbf{\ref{sec:tracking_mechanisms_cached_based}}\mbox{~~~~} & \textbf{Cache-based} &\\
\hline
\textbf{\textit{A}}\mbox{~~} & \textbf{\textit{Web cache} }&  \\
\hline
1 & Embedding identifiers in cached documents & Block JavaScript execution\\
\hline
2 & Loading performance tests & Partly by blocking JavaScript execution, but server-side measurements are still possible\\
\hline
3 & ETags and Last-Modified headers & Web proxies able to remove or rewrite the headers \\
\hline
\textbf{\textit{B}}\mbox{~~} & \textbf{\textit{DNS lookups}} & Block  JavaScript execution \\
\hline
\textbf{\textit{C}}\mbox{~~} & \textbf{\textit{Operational caches}} &  \\
\hline
1 & HTTP 301 redirect cache & No defense known; tracking can be minimized by frequent clearing of browser caches\\
\hline
2 & HTTP authentication cache & No defense known; tracking can be minimized by frequent clearing of browser caches \\
\hline
3 & HTTP Strict Transport Security cache & No defense known; tracking can be minimized by frequent clearing of browser caches \\
\hline
4 & TLS Session Resumption cache and TLS Session IDs & Use Tor Browser \\
\hline
\hline
\textbf{\ref{sec:fingerprinting}}\mbox{~~~~} & \textbf{Fingerprinting} &\\
\hline
\textit{A}\mbox{~~} & \textit{Network and location fingerprinting} & Tor, VPNs, anonymous web proxies, clearing of browser web cache \\
\hline
\textit{B}\mbox{~~} & \textit{Device fingerprinting} & Partly by blocking JavaScript and Flash execution, Tor, VPNs, anonymous web proxies, clearing of browser web cache \\
\hline
\textit{C}\mbox{~~} & \textit{Operating System instance fingerprinting} & Block JavaScript, Flash, Java, and ActiveX execution \\
\hline
\textit{D}\mbox{~~} & \textit{Browser version fingerprinting} & Block JavaScript execution \\
\hline
\textit{E}\mbox{~~} & \textit{Browser instance fingerprinting using canvas} & Block JavaScript execution, use Tor Browser \\
\hline
\textit{F}\mbox{~~} & \textit{Browser instance fingerprinting using web browsing history} & Partly by blocking JavaScript execution, but server-side measurements are still possible \\
\hline
\textit{G}\mbox{~~} & \textit{Other browser instance fingerprinting methods} & Partly by blocking JavaScript and Flash execution \\
\hline
\hline
\textbf{\ref{sec:tracking_mechanisms_sophisticated}}\mbox{~~~~} & \textbf{Other tracking mechanisms} &\\
\hline
\textit{A}\mbox{~~} & \textit{Headers attached to outgoing HTTP requests} & Tor, VPNs, or remote web proxies configured to remove the additional HTTP headers\\
\hline
\textit{B}\mbox{~~} & \textit{Using telephone metadata} & Install mobile applications only from known and trusted sources \\
\hline
\textit{C}\mbox{~~} & \textit{Timing attacks} & Block JavaScript execution\\
\hline
\textit{D}\mbox{~~} & \textit{Using unconscious collaboration of the user} & No defense known; tracking can be minimized by disabling JavaScript and Flash, however, most techniques do not use any of them \\
\hline
\textit{E}\mbox{~~} & \textit{Clickjacking} & No defense known; tracking can be minimized by disabling JavaScript and Flash \\
\hline
\textit{F}\mbox{~~} & \textit{Evercookies (supercookies)} & Partly by disallowing cookies, Flash cookies, blocking JavaScript, Flash, Java, and Silverlight execution, disabling the userData storage in IE, and frequent clearing of all browser caches \\
\hline
\end{tabular}
\label{tab:defense_strategies}
}
\end{table*}


\begin{table*}[t]
\caption{Tracking defense tools}
\centering
\resizebox{2\columnwidth}{!}{
\tiny
\begin{tabular}{|r|m{2cm}|m{3cm}|m{5cm}|}
\hline
\textbf{Section} & \textbf{Name} &  \textbf{Type} & \textbf{Description} \\
\hline
\hline
\textbf{\ref{sec:general_tracking_auditing_and_defense_techniques}}\mbox{~~} & & & \\
\hline
\textit{A} & Microsoft Tracking Protection List & Part of IE & Blocks third-party tracking based on a blacklist. \\
\hline
\textit{A} & Privacy Badger & Browser extension for Google Chrome and Mozilla Firefox & Blocks third-party websites which appear to track the user based on the number of first-party websites embedding the third party. Replaces social network buttons, which blocks the social networks from tracking the user. At the same time, the button functionality is intact. \\
\hline
\textit{A} & Request-Policy & Browser extension for Mozilla Firefox & Blocks all third-party requests by default, by the user can maintain a whitelist. \\
\hline
\textit{A} & Adblock Plus & Browser extension for Mozilla Firefox, Google Chrome, Internet Explorer, Opera, and Safari & Blocks all unwanted ads by default, e.g., video ads on YouTube, Facebook ads, flashy banners, pop-ups, and pop-unders. Adblock Plus is also able to identify acceptable ads. It can be configured to block domains known to spread malware or to disable all known tracking. It also allows to remove all social media buttons from every website. \\
\hline
\textit{B} & Zend2.com, KPROXY, etc. & External anonymous proxy servers & Offer anonymization services, which rely among others on hiding the IP address of the user or/and removing cookies from HTTP requests. Some of them allow bypassing censorship as the user connects with the proxy server instead of with the banned remote website. \\
\hline
\textit{B} & SecurityKISS, CyberGhost, USA IP PPTP/L2TP/OpenVPN VPN service, VPNReactor.com, etc. & Virtual Private Networks (VPNs) & Extend the private network to a remote destination through a secure tunnel. VPN nodes are logically hosts in a single network sharing the same subnet, regardless of where the nodes are physically located. \\
\hline
\textit{B} & Tor & Set of standalone tools for Window Linux, and Mac OS X & Protects the user's IP address by routing the traffic through a chain of Tor relays and, finally, releasing it to the global Internet by the Tor exit relay. Tor Browser is a self-sufficient package for browsing the web anonymously, which contains and embedded Tor client and a browser that disables all possible known kinds of tracking techniques. \\
\hline
\textit{C} & Privoxy & Standalone HTTP and SSL proxy server for Windows, Linux, OpenWrt, DD-WRT, Mac OS X, OS/2, AmigaOS, BeOS, and most Unix distributions & Makes it able to alter, filter, or strip the information requested by the user. Blocks advertisements, banners, popups, and filters traffic, abusive content or cookies. HTTP headers can be modified on the fly. \\
\hline
\textit{I} & NoScript & Browser extension for Mozilla Firefox & Allows JavaScript, Java, Flash, and other plugins to be executed only by trusted web sites of the user's choice. \\
\hline
\textit{I} & Flashblock & Browser extension for Mozilla Firefox & Blocks all Flash content from loading leaving instead placeholders on the webpage. The user can explicitly request loading the specific Flash object. \\
\hline
\textit{K} & Vanish & Browser extension for Mozilla Firefox & Creates self-destroying file systems based on DHT networks (e.g., Vuze, Kademlia, OpenDHT), which can be used for storing sensitive data encapsulated on a website or attached to e-mails send by GMail. \\
\hline
\textit{M} & Disconnect & Standalone commercial tool for Windows, Mac OS X, Android, and IOS & Routes all the traffic through and encrypted tunnel to many selectable destinations, making the user able to avoid censorship and data wiretapping. Blocks around 5000 of trackers, sources of malware and identity theft. Anonymizes search queries made in web browsers. Provides the graphical visualization of tracking entities. \\
\hline
\textit{M} & Meddle & Standalone tool for Android and IOS & Blocks, shapes, filters, or modifies the traffic from mobile devices. Supports ad blocking, tunneling all the traffic through VPN, and provides the visibility of all connections established by mobile phones. The users are informed which personal information is accessed by the particular applications and where the information is sent. They are able to decide which information should be blocked from being sent over the network or how the sent information should be altered. \\
\hline
\end{tabular}
\label{tab:tracking_defense_tools}
}
\end{table*}


\begin{table*}[t]
\caption{Other privacy enhancement techniques}
\centering
\resizebox{2\columnwidth}{!}{
\tiny
\begin{tabular}{|r|m{3cm}|m{5cm}|}
\hline
\textbf{Section} & \textbf{Technique} & \textbf{Description} \\ 
\hline
\hline
\textbf{\ref{sec:general_tracking_auditing_and_defense_techniques}}\mbox{~~} & & \\
\hline
\textit{D} & Opt-out cookies & Unpopular (used by less than 1\% of web browsers) and inconvenient method: \newline - for every third party, a new opt-out cookie must be set \newline - limited lifetime (must be periodically manually renewed) \newline - opt-out cookies are deleted while cleaning all browser cookies \newline Finally, they usually do not stop tracking itself, but only displaying advertisements based on tracking. \\
\hline
\textit{E} & Do Not Track (DNT) browser setting & Although adopted by all main browsers, it does not work in practice as it is supported by a low number of trackers. \\
\hline
\textit{F} & Using privacy-focused search engines (e.g., DuckDuckGo, Startpage, Ixquick) & Contrary to the standard search engines (e.g., Google), they neither collect nor pass the searched terms to the destination websites. \\
\hline
\textit{G} & Private browsing mode & In Firefox, Chrome, and Internet Explorer, it creates a new clean environment where the previously accumulated data (e.g., browsing history, cache, cookies, persistent storages) are not accessible. However, Safari makes the cookies, history, and HTML5 accessible in the private browsing mode, thus, it does not protect the user from being tracked. \\
\hline
\textit{H} & Clearing the browser cache and history & Protects the user from being tracked by cache-based and storage-based mechanisms on the condition that the user does not authenticate in web services (e.g., Facebook) which extends tracking beyond this limitation. \\
\hline
\textit{J} & Using e-mail aliases (offered by, e.g., Inbox Alias, 33Mail, Jeetable, Mailexpire, TrashMail, Guerrilla Mail) & Prevents tracking by e-mail aliases provided on different Internet websites. Decreases the number of spam messages, which also prevents tracking by entities of questionable credibility. \\
\hline
\end{tabular}
\label{tab:other_enhancement_techniques}
}
\end{table*}


\subsection{Blocking advertisement services}

So far, third-party tracking was defended mainly by using blacklists, as Microsoft Tracking Protection List (TPL), which is a part of the newest Internet Explorer \cite{IETrackingProtectionListWeb}. According to \cite{li2014trackadvisor}, this list is characterized by precision of 96.3\,\% and recall of 72.2\,\%. It is possible to avoid third-party tracking by blocking connections to all third-party websites, however, only 37\,\% of the requests to third-party websites are directed to a tracker.

TPLs are designed to block only third-party content. When the user directly navigates to a website from a TPL, it can be accessed without any problems, as it is seen as a first-party website. This can be used by scripts to load the third-party content as the first-party content. The trick is to redirect a user by a JavaScript to the third-party domain, from which he will be after a quick time redirected back to the first-party website. It is also possible to use popup windows for this purpose -- the loaded third-party content will also appear as coming from the first party \cite{blackhat2012ImplementingWebTrackingWeb}.

Facebook admits that while we browse a website that contains the \emph{Like} button, the information about our Facebook identifier is sent to the server in order to show us who of our friends like the page as well \cite{FacebookLikeOfficialWeb}. At the same time, some additional browser-related information is sent to Facebook together with the address of the website we browse \cite{FacebookLikeOfficialWeb}. The information collected by Atlas \cite{AtlasSolutionsWeb} (the underlying tracking mechanism) from various sources (including the Facebook profiles) are used to sell advertisements by Facebook on properties which are owned by other entities \cite{FacebookWillUseFacebookDataWeb}. ShareMeNot \cite{ShareMeNotWeb}, incorporated from July 2014 into Privacy Badger \cite{PrivacyBadgerWeb}, is able to detect and selectively block third-party buttons (e.g., Google \emph{+1} or Facebook \emph{Like}) from tracking the user unless the user clicks them. That way, the functionality of the websites in not limited, but the user's privacy is protected from being tracked by the buttons on all the visited websites. ShareMeNot uses multiple approaches to achieve this goal: provides replacement buttons instead of the original ones, disables requests to the trackers, or optionally removes cookies from request made while real buttons are loaded. When a user chooses to click on a button, ShareMeNot allows the button provider to identify the user, so the initial functionality of the button is preserved.

A very similar addon for a Firefox web browser is RequestPolicy \cite{RequestPolicyWeb}. By default, it blocks all the cross-site requests (requests to the third parties), but the user can maintain a whitelist.

Adblock Plus is an open source popular browser extension created in 2006, which reached from that time more than 50 million users. It blocks all unwanted ads by default, e.g., video ads on YouTube, Facebook ads, flashy banners, pop-ups, and pop-unders. Adblock Plus is also able to identify acceptable ads. It can be configured to block domains known to spread malware or to disable all known tracking. It also allows to remove all social media buttons from every website \cite{AdblockPlusWeb}. In December 2014, the French division of the Interactive Advertising Bureau (IAB) \cite{IABAboutTheIABWeb} decided to fill a complaint to the court against the software allowing hiding advertisements, especially Adblock, which is the market leader.

\subsection{Hiding the IP address}

The most popular method to hide the IP address from the remote site is to use anonymous proxy servers, Virtual Private Networks (VPNs), or Tor \cite{schmucker2011web}.

Proxy servers are intermediary entities in communication between two parties. Some of the proxy servers offer anonymization services, which rely among others on hiding the IP address of the user or/and removing cookies from HTTP requests. At the same time, some of the proxy services allow the user to bypass censorship as the user connects with the proxy server instead of with the banned remote website. There are many web-based anonymous proxy services offered by free, e.g., Zend2.com \cite{Zend2comWeb} and KPROXY \cite{KPROXYWeb}.

VPNs are designed to extend the private network to a remote destination through a secure tunnel. VPN nodes are logically hosts in a single network sharing the same subnet, regardless of where the nodes are physically located. The tunnel connecting the particular parts of a VPN is usually secured by one of the following protocols: Point-to-Point Tunneling Protocol (PPTP), Layer 2 Tunneling Protocol (L2TP), Internet Protocol Security (IPsec), Secure Sockets Layer (SSL), or OpenVPN. SecurityKISS \cite{SecurityKISSWeb}, CyberGhost \cite{CyberGhostWeb}, USA IP PPTP/L2TP/OpenVPN VPN service \cite{USAIPWeb}, and VPNReactor.com L2TP/PPTP/OpenVPN VPN service \cite{VPNReactorWeb} are examples of widely-known VPN services.

Tor bundles \cite{TorProjectAnonymityOnlineWeb} consist of the onion routing software and Tor Browser, which is a self-sufficient package for browsing the web anonymously. The users can select to install only the core onion routing package and use it together with their own web browsers, although it is not recommended, as Tor Browser protects the user's privacy by disabling all possible known kinds of tracking techniques. The traffic is routed through a chain of Tor relays and, finally, leaves the network by the Tor exit relay, which protects the real user's IP address from being known to the remote parties. Tor packages for running Tor non-exit and exit relays are also accessible on the developers website.

\subsection{Modification of data sent over the network}

Privoxy \cite{PrivoxyHomePageWeb} is an HTTP and SSL proxy server, which is able to alter, filter, or strip the information requested by the user. The main functionality of this proxy is concentrated on blocking advertisements, banners, and popups, and filtering tracking or abusive content or cookies. Privoxy can also modify the HTTP headers on the fly.

\subsection{Opt-out cookies}

Users can opt-out from behavioral advertising using opt-out cookies. However, according to \cite{WebTrackingUserPrivacyWorkshop2011Web}, less than 1\,\% of web browsers are reported to use them. There are several grounds for such a state of affairs. At first, for every third party, a new tracking cookie must be set. At second, these cookies have a limited lifetime, so they must be periodically manually renewed. At third, the cookies are lost when the user cleans the cookies from his web browser \cite{mayer2012third}. At fourth, the opt-out does not block the tracking itself, but only providing the advertisements based on the tracking \cite{TrackingTrackersSelfHelpToolsWeb}. 

\subsection{Do Not Track}

Do Not Track (DNT) \cite{TrackingPreferenceExpressionDNTWeb} is a new type of HTTP header field, which can be appended to outgoing requests to inform the service that the user does not want to be tracked. DNT also defines an HTML DOM property, which is accessible to scripts running on the website, and APIs to register site-specific exceptions granted by the user. The websites can include the \emph{Tk} header in the HTTP response to communicate whether and how they honor a received preference \cite{TrackingPreferenceExpressionDNTWeb}. DNT was widely adopted by all the main web browsers. It was aimed at becoming the standard technique to opt-out web tracking by different services, however, it requires the tracker compliance and there is no technical or lawful possibility to enforce this request \cite{Roesner2012DDA22282982228315}.

\subsection{Using privacy-focused search engines}

Privacy-focused search engines (e.g., DuckDuckGo \cite{DuckDuckGoWeb}, Startpage Search Engine \cite{StartpageSearchEngineWeb}, and Ixquick Search Engine \cite{IxquickSearchEngineWeb}) claim that they do not collect any private information. Normally, when following the link from a web browser, the \emph{Referer} field in the HTTP header contains also the searched phrases. However, the privacy-focused search engines use some techniques to prevent forwarding the search results to the linked website, mainly by obfuscating the \emph{Referer} field in the HTTP header, or by using POST instead of GET requests.

\subsection{Private browsing mode}

All the main browsers developed a private browsing mode in order to provide the user more anonymity. A comprehensive comparison of the implementations of private browsing by various modern browsers is shown in \cite{bursztein2010analysis}. All the main browsers (Firefox, Safari, Chrome, and Internet Explorer) provide quite good defense against the local attacker, who is, therefore, not able to determine which websites were visited by the user on the condition that the attacker did not have any prior access to the computer and the user did not bookmark any websites or saved any files on his hard drive. Regarding the web attacker, the situation is quite different. Safari, contrary to the other browsers, ignores this threat by making its earlier cookies, history, and HTML5 storages accessible in the private browsing mode. Another threat are the fingerprinting techniques, for example, used by fingerprintJS \cite{ValveFingerprintjsGitHub}, which return the same values in the normal and private modes. Table~\ref{tab:private_bowsing_mode} shows how using the private browsing mode influences the different tracking techniques.


\begin{table*}[p]
\caption{Private browsing mode}
\centering
\resizebox{2\columnwidth}{!}{
\tiny
\begin{tabular}{|r|l|m{5cm}|}
\hline
\textbf{Section} & \textbf{Tracking mechanisms} & \textbf{Tracking a normal-mode user identity in a private browsing mode} \\ 
\hline
\hline
\textbf{\ref{sec:tracking_mechanisms_session_only}}\mbox{~~~~} & \textbf{Session-only} &\\
\hline
\textit{A}\mbox{~~} & \textit{Session identifiers stored in hidden fields} & Not applicable\\
\hline
\textit{B}\mbox{~~} & \textit{Explicit web-form authentication} & Not applicable \\
\hline
\textit{C}\mbox{~~} & \textit{window.name DOM property} & Not applicable \\
\hline
\hline
\textbf{\ref{sec:tracking_mechanisms_client_storage}}\mbox{~~~~} & \textbf{Storage-based} &\\
\hline
\textit{A}\mbox{~~} & \textit{HTTP cookies} & Yes - in Safari \newline No - in Chrome, Firefox and IE \\
\hline
\textit{B}\mbox{~~} & \textit{Flash cookies} & Yes - in Safari \newline No - in Chrome, Firefox and IE \\
\hline
\textit{B}\mbox{~~} & \textit{Java JNLP PersistenceService} & Yes \\
\hline
\textit{C}\mbox{~~} & \textit{Flash LocalConnection object} & Yes \\
\hline
\textit{D}\mbox{~~} & \textit{Silverlight Isolated Storage} & Yes - in Safari \newline No - in Chrome, Firefox and IE \\
\hline
\textit{E}\mbox{~~} & \textit{HTML5 Global, Local, and Session Storage} & Yes - in Safari \newline No - in Chrome, Firefox and IE \\
\hline
\textit{F}\mbox{~~} & \textit{Web SQL Database and HTML5 IndexedDB} & Yes - in Safari \newline No - in Chrome, Firefox and IE \\
\hline
\textit{G}\mbox{~~} & \textit{Internet Explorer userData storage} & No \\
\hline
\hline
\textbf{\ref{sec:tracking_mechanisms_cached_based}}\mbox{~~~~} & \textbf{Cache-based} &\\
\hline
\textbf{\textit{A}}\mbox{~~} & \textbf{\textit{Web cache} }&  \\
\hline
1 & Embedding identifiers in cached documents & Yes - in Safari \newline No - in Chrome, Firefox and IE \\
\hline
2 & Loading performance tests & Yes - in Safari \newline No - in Chrome, Firefox and IE\\
\hline
3 & ETags and Last-Modified headers & Yes - in Safari \newline No - in Chrome, Firefox and IE \\
\hline
\textbf{\textit{B}}\mbox{~~} & \textbf{\textit{DNS lookups}} & Yes \\
\hline
\textbf{\textit{C}}\mbox{~~} & \textbf{\textit{Operational caches}} &  \\
\hline
1 & HTTP 301 redirect cache & Yes - in Safari \newline No - in Chrome, Firefox and IE \\
\hline
2 & HTTP authentication cache & Yes - in Safari \newline No - in Chrome, Firefox and IE \\
\hline
3 & HTTP Strict Transport Security cache & Yes - in Safari \newline No - in Chrome, Firefox and IE \\
\hline
4 & TLS Session Resumption cache and TLS Session IDs & Yes - in Safari \newline No - in Chrome, Firefox and IE \\
\hline
\hline
\textbf{\ref{sec:fingerprinting}}\mbox{~~~~} & \textbf{Fingerprinting} &\\
\hline
\textit{A}\mbox{~~} & \textit{Network and location fingerprinting} & Yes \\
\hline
\textit{B}\mbox{~~} & \textit{Device fingerprinting} & Yes \\
\hline
\textit{C}\mbox{~~} & \textit{Operating System instance fingerprinting} & Yes \\
\hline
\textit{D}\mbox{~~} & \textit{Browser version fingerprinting} & Yes \\
\hline
\textit{E}\mbox{~~} & \textit{Browser instance fingerprinting using canvas} & Yes \\
\hline
\textit{F}\mbox{~~} & \textit{Browser instance fingerprinting using web browsing history} & Yes - in Safari \newline No - in Chrome, Firefox and IE \\
\hline
\textit{G}\mbox{~~} & \textit{Other browser instance fingerprinting methods} & Yes \\
\hline
\hline
\textbf{\ref{sec:tracking_mechanisms_sophisticated}}\mbox{~~~~} & \textbf{Other tracking mechanisms} &\\
\hline
\textit{A}\mbox{~~} & \textit{Headers attached to outgoing HTTP requests} & Yes \\
\hline
\textit{B}\mbox{~~} & \textit{Using telephone metadata} & Not applicable \\
\hline
\textit{C}\mbox{~~} & \textit{Timing attacks} & Yes - in Safari \newline No - in Chrome, Firefox and IE \\
\hline
\textit{D}\mbox{~~} & \textit{Using unconscious collaboration of the user} & Yes - in Safari \newline No - in Chrome, Firefox and IE \\
\hline
\textit{E}\mbox{~~} & \textit{Clickjacking} & Not applicable \\
\hline
\textit{F}\mbox{~~} & \textit{Evercookies (supercookies)} & Yes - in Safari \newline No - in Chrome, Firefox and IE \\
\hline
\end{tabular}
\label{tab:private_bowsing_mode}
}
\end{table*}


\subsection{Clearing the browser cache and history}

This method, on the condition that the user properly cleans every storage in which the browser and its plugins can keep information, behaves almost in the same way as private browsing. The only exception is Safari, which in the private browsing mode continuously serves cookies and history from the earlier browsing activities.

\subsection{Execution blocking}

There are several browser extensions, which are able to block the execution of JavaScript (e.g., NoScript \cite{NoScriptSecuritySuiteFirefoxWeb}), Flash (e.g., Flashblock \cite{FlashblockFirefoxWeb}), and other script content. However, other tracking methods are not impacted by using these extensions.

\subsection{E-mail aliases}

One of the most frequently revealed on the websites sensitive information is the e-mail address. If the e-mail address is publicly accessible (for example, revealed in the user's public profile), it allows easy association of different accounts to the same user. Furthermore, publicly available e-mail addresses are harvested by spam bots and used as the target for unsolicited advertising and, thus, further tracking of the user. It is why the protection of e-mail addresses is very important. However, it is not practical to create multiple e-mail accounts, which need to be frequently checked. A solution for this problem is to use e-mail aliases, from which the incoming messages are automatically forwarded to the user's real account. Many providers (e.g., Inbox Alias \cite{InboxAliasWeb}, 33Mail \cite{33MailcomWeb}, Jetable \cite{JetableorgWeb}, Mailexpire \cite{MailexpireWeb}, TrashMail \cite{TrashMailWeb}, and Guerrilla Mail \cite{GuerrillaMailWeb}) offer free e-mail aliases, which lifetime can be configured from 5 minutes, through 1 month, to being infinite. They can also be manually deleted if not needed anymore. That prevents spam and protects the users' privacy, as the e-mail aliases can be safely provided to any website which requires a working e-mail address, or used to be displayed in a public web profile.

\subsection{Self-destroying file systems}

Sometimes, it is desired to place on a web or send by e-mail some content, which has a limited lifetime. That especially concerns sensitive data that should not be accessible in any way after the time specified by the owner, which prevents extracting the sensitive information at a later time from the collected data. Self-destroying file systems are the response to this need.

In \cite{geambasu2009vanish}, the authors presented Vanish -- a novel self-destroying file system, which can be used for sharing sensitive data. This system is based on distributed DHT networks (e.g., Vuze, Kademlia, OpenDHT), which are used to store the encapsulated user's data. The authors implemented prototypes of Vanish clients as a Firefox plugin for sending and reading destroyable e-mails from GMail, and a browser extension, which allows to integrate encapsulated destroyable elements on a website.

\subsection{Discovering how tracking works}

A summary of available traffic auditing tools is shown in Table~\ref{tab:traffic_auditing_tools}. Regardless of the mechanism, tracking can be automatically discovered by correlation between the sources of information associated with the user (as the list of visited websites or received e-mails) and the advertisements presented to him. XRay -- an automatic system for discovering these correlations -- is shown in \cite{lecuyer2014xray}. The system was tested and evaluated separately on GMail, Amazon, and YouTube. Additionally, the authors assessed its precision and recall across services (YouTube to GMail). XRay was shown to be able to correlate GMail advertisements, products suggested by Amazon, and videos recommended on YouTube to e-mails in GMail, Amazon wishlists, and previously watched videos on YouTube. The correlation is based on creating a number of shadow accounts, which are used to perform various activities and to compare the produced output. For example, if an advertisement is seen on multiple accounts which contain the same e-mail while this ad is never shown on accounts lacking this particular e-mail, it is highly probable that this ad is directly associated with the content of this particular e-mail. The authors showed that XRay needs only a logarithmic number of accounts in the number of data inputs. The experiments showed that all the e-mails related to various diseases (except AIDS) result in the advertisements specific to the topic. For example, e-mails related to depression cause numerous \emph{shamanic healing} ads, e-mails related to Alzheimer disease cause appearance of assisted living services ads, while cancer-related e-mails cause ads related to fight-cancer campaigns. Other topics, which highly correspond to the high number of associated ads are pregnancy, divorce, race, and homosexuality. It was shown that e-mails related to pregnancy result in showing baby-related ads, while e-mails related to homosexuality result in advertisements of gay-friendly hotels.


\begin{table*}[t]
\caption{Tracking auditing tools}
\centering
\resizebox{2\columnwidth}{!}{
\tiny
\begin{tabular}{|r|m{1.2cm}|m{3cm}|m{5cm}|}
\hline
\textbf{Section} & \textbf{Name} &  \textbf{Type} & \textbf{Description} \\ 
\hline
\hline
\textbf{\ref{sec:general_tracking_auditing_and_defense_techniques}}\mbox{~~} & & \\
\hline
\textit{L} & XRay & Browser extension for Google Chrome & Discovers the correlations between the sources of information associated with the user (e.g., the list of visited websites or received e-mails) and the advertisements presented to him. \\
\hline
\textit{L} & TaintDroid & Android application & Tracks in real time which private data (e.g., user's location, device identifier, SIM card and phone numbers) are leaked out of the mobile device and which applications are responsible for the leaks. \\
\hline
\textit{L} & Lightbeam & Browser extension for Mozilla Firefox & Provides interactive visualization and shows the relationships between tracking third parties as well as all the requests made to the first and third party sites. \\
\hline
\textit{L} & FPDetective & Analytical framework shipped as a VirtualBox virtual machine & Identifies and analyzes web-based fingerprinting by JavaScript and Flash. \\
\hline
\textit{L} & MindYourPrivacy & Standalone tool for Linux & Uses Deep Packet Inspection techniques to analyze and visualize collectors of private information based on HTTP Referers. \\
\hline
\textit{L} & Sheriff & Browser extension for Google Chrome & Allows the user to select the price of a particular product on the website of the web store and check what is the price of this product if the website is accessed by various web browsers and from different geographical locations. \\
\hline
\textit{L} & Disconnect free & Standalone tool for Windows, Mac OS X, Android, and iOS & Visualizes third-party trackers on Internet websites. \\
\hline
\end{tabular}
\label{tab:traffic_auditing_tools}
}
\end{table*}  


TaintDroid \cite{enck2014taintdroid} is an extension to Android, which tracks in real time how different applications use our private data, especially, which data is leaked out of the mobile device. It labels data coming from privacy-sensitive sources and, thus, it can monitor when such data leave the system. The authors evaluated the accuracy of TaintDroid by installing 30 random popular applications from Google Play store. The test revealed that 15 out of the 30 applications reported users' locations to remote advertising servers. Seven of the examined applications collected the device identifier, while some of them also collected the SIM card serial number and the phone number.

Lightbeam for Firefox \cite{LightbeamForFirefoxWeb} enables the user to see the requests made to the first and third party sites. By providing interactive visualizations, Lightbeam shows the relationships between these parties.

FPDetective \cite{acar2013fpdetective} is a framework for identifying and analyzing of web-based fingerprinting. It consists of a crawler, a parser, a proxy, a decompiler, and a central database. The crawler is based on two approaches: JavaScript fingerprinting data is collected by PhantomJS driven by CasperJS, while Flash fingerprinting data is collected by Chromium driven by Selenium WebDriver. WebKit (the engine used by both browsers) was modified to intercept and log access to various properties using API calls, which can be used for fingerprinting. All the traffic was passed by \emph{mitmproxy}, where \emph{mitmdump} logged all the HTTP traffic and \emph{libmproxy} parsed and extracted all Flash files, which were decompiled to their original ActionScript source code using \emph{JPEXS Free Flash Decompiler}. A Flash file was considered to be fingerprinting when it enumerated system fonts, collected information about the devices, and sent the information to a remote server by socket connections (\emph{sendAndLoad}, \emph{URLLoader} calls) or JavaScript asynchronous calls (\emph{ExternalInterface.Call}, \emph{addCallback}). A JavaScript file was considered to be fingerprinting when it loaded more than 30 system fonts, enumerated plugins or mime types, detected screen or browser properties, and sent the data back to the server.

MindYourPrivacy \cite{takano2014mindyourprivacy} is a tool for the visualization of collectors of private information, which is based on Deep Packet Inspection techniques. Thanks to that, the authors achieved browser and device independence. The results of the analysis are provided through a web-based user-friendly interface. This tool provide a wide range of identification techniques. At first, it analyzes and visualizes the graph of HTTP \emph{Referers}. The tracking domains are aggregated as either second or third-level domains. The system consists of an application-level packet analyzer, NoSQL DB for data storage, tracking analyzer, results generator, and an HTTP server. The main outputs of the system are: most-referred URLs, sites referring most-referred URLs, and the graph of the HTTP referrers.

\$heriff \cite{SheriffDetectingPriceDiscriminationWeb} is a service for discovering price discrimination over the Internet. It is available as an extension for Google Chrome. \$heriff allows the user to select the price of a particular product on the website of the online store and check what is the price of this product if the website is accessed by various web browsers and from different geographical locations. For that purpose, a query is issued to the \$heriff server, which issues further queries with different \emph{User-Agent} field and from different geographical locations using the PlanetLabs infrastructure.

\subsection{Combined tools}

Disconnect \cite{DisconnectOnlinePrivacySecurityWeb} is a commercial tool combining multiple approaches shown above. It routes all the traffic through an encrypted tunnel to many selectable destinations, making the user able to avoid the censorship and data wiretapping. The VPN blocks around 5000 of trackers, sources of malware and identity theft. The search queries made in web browsers are anonymized, so the browser cannot associate them with the particular user. This tool also provides the graphical visualization of tracking entities. A free version of Disconnect exists, however, its capabilities are limited to the visualization of trackers.

Meddle \cite{MeddleWeb}, described in detail in a technical report \cite{meddle2013TechReport}, is a tool for Android and IOS designed to block, shape, filter, or modify the traffic from mobile devices. It supports ad blocking, tunneling all the traffic through VPN, and provides the visibility of all connections established by mobile phones. The users are informed which personal information is accessed by the particular applications and where the information is sent. They are able to decide which information should be blocked from being sent over the network or how the sent information should be altered. All the modifications are made by a chain of middle boxes.

\section{The future of tracking}
\label{sec:the_future_of_tracking}

\subsection{The Future of the Cookie Working Group}

The Future of the Cookie Working Group \cite{TheFutureOftheCookieWorkingGroupIABWeb} of the Interactive Advertising Bureau (IAB) \cite{IABAboutTheIABWeb} is designing a new technology to track users across multiple sessions and devices. The group consists of 90 member companies, including [x+1], Adobe, Amazon, AOL, eBay, Google, PayPal, Quantcast, Verizon, and Yahoo \cite{TheFutureOftheCookieWorkingGroupIABWeb}. The IAB, the host association, consists of more than 650 leading media and technology companies that are responsible for selling, delivering, and optimizing digital advertising, or marketing campaigns. It claims to account for 86\,\% of online advertising in the US \cite{IABAboutTheIABWeb}.

In the whitepaper published in January 2011 \cite{PrivacyTrackingPostCookieWorldWhitepaper}, the group mentions that the current tracking techniques are not sufficient. Third-party cookies are often disabled by privacy-concerned users (and by Firefox, by default). The users are considering legislation approaches to regulate current tracking policies. All of these are causing loss of revenue to the advertising services. The authors claim that cookies have too less power: there can be accessed only by the same domain as set it up, they are specific for a device, a login to the device, and each installed web browser. The report indicates that the web services cannot use the information collected from a user from his laptop (as the location set for the weather forecast) in order to provide the same experience on the mobile device, unless the users logs in (which, in our opinion, should be the only way the user has its webspace personalized). The aim of this new tracking project is to provide full transparency of the consumer. In order to achieve that, five solution classes were proposed:

\begin{itemize}

 \item Device-inferred, using an identifier created by using statistical algorithms based on the information about the device, browser, application, or the operating system. The identifier should be tied to the specific device and known to multiple third parties if they use the same statistical algorithm. However, the authors noticed that it is not easy to keep the identifier stable, as users change their network or IP addresses, update their devices or operating systems, and change their browser add-ons.

 \item Client-generated, using a static identifier, which is passed to all the third parties together with the browser, application, and operating system settings, which can be used to customize the advertising experience. The methodology is similar to Apple’s AdID and identifiers used by Google and Microsoft.

 \item Network-inserted, using an identifier assigned by a third-party intermediary server on the path between the user's device and the publishers' servers. That can be achieved by the use of content distribution networks, Wi-Fi or wireless proxy servers, and ISPs. All the users' devices could be associated with a single identifier if the devices are registered within the same ISP as belonging to the same user. A network identifier could also be used across different networks in case of network partnerships.

 \item Server-issued, using cookies.

 \item Cloud-synchronized, using an identifier assigned by a centralized service for all third parties.

\end{itemize}

\subsection{Google AdID}

According to an article from September 2013 \cite{GoogleMayDitchOnlineTracker}, Google is developing AdID, an anonymous identifier, which can be used for advertising and replace third-party cookies. This identifier is intended to be transmitted to the parties which agree to some guidelines giving the users more privacy and control over the tracking. The AdID is planned to be reset every year. Different browsing profiles will be associated with different AdIDs, giving the users possibility to keep their private sessions disconnected from their normal browsing activities. Additionally, the users will be able to block specific advertisers by the control panel in the browsers.

\subsection{Microsoft Device Identifier}

An article from October 2013 \cite{MicrosoftCookieReplacementSpanDesktopMobileXbox} reveals that Microsoft is working on developing its own tracking technology using a device identifier. The technology is designed to span desktop, mobile, and xbox. The permission to use the identifier will be given to Microsoft by the user when accepting the user agreement or terms of service.

\subsection{Data Transparency Lab}

Data Transparency Lab (DTL) \cite{DataTransparencyLabWeb} is a community, which aims at bringing clarity to the flow and usage of personal data online and in exploring ways towards a transparent and respectful data trade in the future. DTL consists of universities, businesses, and other institutions, which discover new tracking methods, analyze their impact, and design tools, to make the users able to control the use of their personal information on the web.

\section{Conclusion}
\label{sec:conclusion}

The aim of this survey was to familiarize the reader with the different tracking mechanisms he, or she, encounters when browsing the web on a regular basis. The presented mechanisms have different scopes, coverages, and purposes. We show that the methods evolved over time from being privacy-friendly (as HTTP cookies, which can be easily removed by the users in case if not desired by the user) to be more and more invasive. Some of the recent techniques (e.g., fingerprinting, or using special identifiers inserted by ISPs) are not easy to avoid by regular users. Even worse, the currently designed methods are potentially even more harmful. However, during the last few years, it is observed a subsequent substantial growth of tools and services that can be used to protect the users' privacy. The potential defense approaches were described and evaluated against different tracking methods. We also showed a number of tools that can be used to discover which private data and by which application are leaking from a computer or a mobile phone.

\section*{Acknowledgment}

This work was funded by Spanish Ministry of Economy and Competitiveness under contract EUIN2013-51199 (Arquitectura con Conocimiento del Entorno de la Futura Internet project), Spanish Ministry of Science and Innovation under contract TEC2011-27474 (NOMADS project), and by AGAUR (ref. 2014-SGR-1427).

\bibliographystyle{IEEEtran}
\bibliography{bibliography}

\begin{IEEEbiography}[{\includegraphics[width=1in,height=1.25in,clip,keepaspectratio]{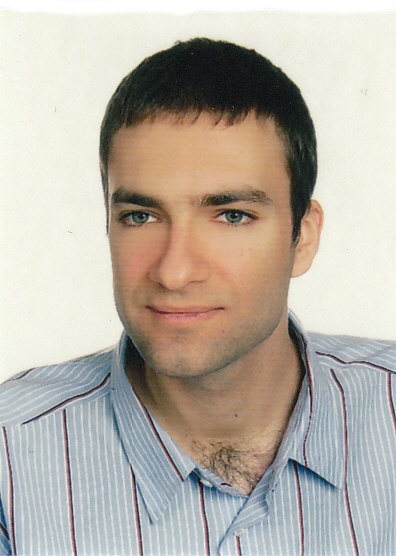}}]{Tomasz Bujlow}
received the M.Sc. and Ph.D. degrees in Computer Science from the Silesian University of Technology in 2008 and Aalborg University in 2014, respectively. He is currently a Postdoctoral Researcher at the Broadband Communications Research Group (CBA) that belongs to the Computer Architecture Department (DAC) at the UPC BarcelonaTech. His research interests are in the field of traffic analysis and network measurements, focusing on network traffic classification. He is a participant on numerous scientific projects related to traffic classification and he holds the Cisco Certified Network Professional (CCNP) certification since 2010. See more: \url{http://tomasz.bujlow.com}.
\end{IEEEbiography}

\begin{IEEEbiography}[{\includegraphics[width=1in,height=1.25in,clip,keepaspectratio]{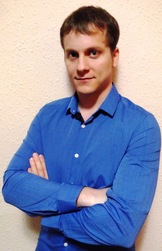}}]{Valentín Carela-Español}
received a B.Sc. degree in Computer Science from the Universitat Politècnica de Catalunya (UPC) in 2007, a M.Sc. degree in Computer Architecture, Networks, and Systems from UPC in 2009, and a Ph.D. degree from UPC in 2014. He is currently a Postdoctoral Researcher at the Broadband Communications Research Group (CBA) that belongs to the Computer Architecture Department (DAC) at the UPC BarcelonaTech. His research interests are in the field of traffic analysis and network measurements, focusing on network traffic classification. His key research area is the study of the identification of applications in network traffic based on Machine Learning and Deep Packet Inspection techniques and the aspects related to the application of those techniques in backbone networks. See more: \url{http://people.ac.upc.es/vcarela}.
\end{IEEEbiography}

\begin{IEEEbiography}[{\includegraphics[width=1in,height=1.25in,clip,keepaspectratio]{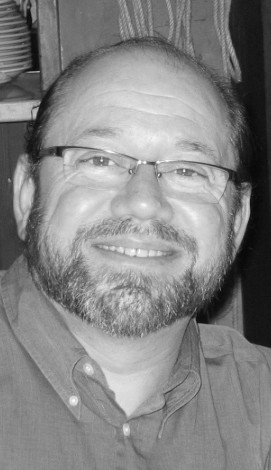}}]{Josep Solé-Pareta}
obtained his M.Sc. degree in Telecom Engineering in 1984, and his Ph.D. in Computer Science in 1991, both from the Universitat Politècnica de Catalunya (UPC). In 1984 he joined the Computer Architecture Department of the UPC. Currently he is Full Professor with this department. He did a Postdoc stage (summers of 1993 and 1994) at the Georgia Institute of Technology. He is co-founder of the UPC-CCABA (\url{http://www.ccaba.upc.edu}). His publications include several book chapters and more than 100 papers in relevant research journals (>25), and refereed international conferences. His current research interests are in Nanonetworking Communications, Traffic Monitoring and Analysis and High Speed and Optical Networking, with emphasis on traffic engineering, traffic characterization, MAC protocols and QoS provisioning. He has participated in many European projects dealing with Computer Networking topics. See more: \url{http://people.ac.upc.es/pareta}.
\end{IEEEbiography}

\begin{IEEEbiography}[{\includegraphics[width=1in,height=1.25in,clip,keepaspectratio]{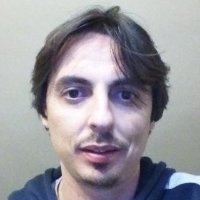}}]{Pere Barlet-Ros}
received the M.Sc. and Ph.D. degrees in Computer Science from the Universitat Politècnica de Catalunya (UPC) in 2003 and 2008, respectively. He is currently an Associate Professor with the Computer Architecture Department of UPC and co-founder of Talaia Networks, a University spin-off that develops innovative network monitoring products. His research interests are in the fields of network monitoring, traffic classification, and anomaly detection. See more: \url{http://people.ac.upc.es/pbarlet}.
\end{IEEEbiography}

\end{document}